\documentclass[11pt]{article}

\usepackage{graphics}
\usepackage{amssymb,amsmath,amsthm, mathtools}

\usepackage{bbm}
\makeatletter
\if@twoside \oddsidemargin -17pt \evensidemargin 00pt
\else \oddsidemargin 00pt \evensidemargin 00pt
\fi
\expandafter\ifx\csname mathrm\endcsname\relax\def\mathrm#1{{\rm #1}}\fi

\@addtoreset{equation}{section}

\makeatother

\def\refeq#1{\mbox{(\ref{#1})}}

\def\refse#1{\mbox{Section~\ref{#1}}}
\def\refses#1{\mbox{Sections~\ref{#1}}}

\def\citere#1{\mbox{Ref.~\cite{#1}}}
\def\citeres#1{\mbox{Refs.~\cite{#1}}}

\def\ie{i.e.\ }
\def\eg{e.g.\ }

\allowdisplaybreaks[1]   

\newcommand{\rb}[1]{\left| #1 \right\rangle}
\renewcommand{\Im}[1]{\mathop{\mathrm{Im}\left[ #1 \right]}\nolimits}
\renewcommand{\Re}[1]{\mathop{\mathrm{Re}\left[ #1 \right]}\nolimits}

\newcommand{\Ree}{\mathop{\mathrm{Re}}\nolimits}
\newcommand{\abs}[1]{\left| #1 \right|}

\renewcommand{\vec}{\mathbf}
\newcommand{\DeltaA}{\Delta_{\mathrm{A}}}
\newcommand{\DeltaR}{\Delta_{\mathrm{R}}}

\newcommand{\DeltaF}{\Delta_{\mathrm{F}}}
\newcommand{\imag}{\mathrm{i}}

\newcommand{\kreis}[1]{\unitlength1ex{\strut\begin{picture}(2.5,2.5)%
\put(0.75,0.75){\circle{2.5}}\put(0.75,0.75){\makebox(0,0){#1}}\end{picture}}}

\makeatletter
\newcommand{\specialnumber}[1]{%
  \def\tagform@##1{\maketag@@@{(\ignorespaces##1\unskip\@@italiccorr#1)}}%
}
\newcommand{\specialeqref}[2]{\begingroup
  \def\tagform@##1{\maketag@@@{(\ignorespaces##1\unskip\@@italiccorr#2)}}%
  \eqref{#1}\endgroup}
\makeatother

\def\draftdate{\relax}
\def\mda{\relax}
\def\mua{\relax}
\def\mla{\relax}
\def\Mda{\relax}
\def\Mua{\relax}
\def\Mla{\relax}
\def\draft{
\def\thtystars{******************************}
\def\sixtystars{\thtystars\thtystars}
\def\lra{\mathop{\mathrm{\leftrightarrow}}\nolimits}
\typeout{}
\typeout{\sixtystars**}
\typeout{* Draft mode!
         For final version remove \protect\draft\space in source file *}
\typeout{\sixtystars**}
\typeout{}
\def\draftdate{\today}
\def\mua{\marginpar[\boldmath\hfil$\uparrow$]%
                   {\boldmath$\uparrow$\hfil}%
                    \typeout{marginpar: $\uparrow$}\ignorespaces}
\def\mda{\marginpar[\boldmath\hfil$\downarrow$]%
                   {\boldmath$\downarrow$\hfil}%
                    \typeout{marginpar: $\downarrow$}\ignorespaces}
\def\mla{\marginpar[\boldmath\hfil$\rightarrow$]%
                   {\boldmath$\leftarrow $\hfil}%
                    \typeout{marginpar: $\lra$}\ignorespaces}
\def\Mua{\marginpar[\boldmath\hfil$\Uparrow$]%
                   {\boldmath$\Uparrow$\hfil}%
                    \typeout{marginpar: $\uparrow$}\ignorespaces}
\def\Mda{\marginpar[\boldmath\hfil$\Downarrow$]%
                   {\boldmath$\Downarrow$\hfil}%
                    \typeout{marginpar: $\downarrow$}\ignorespaces}
\def\Mla{\marginpar[\boldmath\hfil$\Rightarrow$]%
                   {\boldmath$\Leftarrow $\hfil}%
                    \typeout{marginpar: $\lra$}\ignorespaces}
\overfullrule 5pt
\oddsidemargin -15mm
\oddsidemargin -10mm
\marginparwidth 29mm
}

\begin{document}

\thispagestyle{empty}
\def\thefootnote{\fnsymbol{footnote}}
\setcounter{footnote}{1}
\null
\hfill
\\
\vskip 2cm
\begin{center}
  {\Large \boldmath{\bf The Complex-Mass Scheme and Unitarity\\[1ex]
   in Perturbative Quantum Field Theory}
\par} \vskip 2.5em
{\large
{ Ansgar~Denner, Jean-Nicolas~Lang}\\[3ex]
{\normalsize \it
Universit\"at W\"urzburg, 
Institut f\"ur Theoretische Physik und Astrophysik, \\
D-97074 W\"urzburg, Germany}\\[1ex]
}
\par \vskip 1em
\end{center}\par
\vfill \vskip .0cm \vfill {\bf Abstract:} \par


We investigate unitarity within the Complex-Mass Scheme, a convenient
universal scheme for perturbative calculations involving unstable
particles in Quantum Field Theory which guarantees exact gauge
invariance.  Since this scheme requires to introduce complex masses
and complex couplings, the Cutkosky cutting rules, which express
perturbative unitarity in theories of stable particles, are no longer
valid.  We derive corresponding rules for scalar theories with
unstable particles based on Veltman's Largest-Time Equation and
prove unitarity in this framework.

\par
\vskip 1cm
\noindent
April 2015
\par
\null
\setcounter{page}{0}
\clearpage
\def\thefootnote{\arabic{footnote}}
\setcounter{footnote}{0}


\section{Introduction}
With the discovery of the Higgs boson at the Large Hadron Collider nature again
reflects not only the relevance of fundamental principles such as gauge
invariance as they are incorporated in theories like the Standard Model (SM),
but also that unstable particles are as much important as stable ones.  The
majority of the known fundamental particles are unstable, and in physical
observables unstable particles usually play a significant role. 

Precision predictions within perturbative Quantum Field Theories (QFT) are still
a challenging task, especially when unstable particles are involved. Unstable
particles are of non-perturbative nature in the sense that in the usual
leading-order (LO) perturbation theory all particles are stable. As a
consequence near thresholds or resonances observables even diverge in standard
perturbation theory because important contributions are missing.  A proper
treatment requires the inclusion of finite-width effects via a finite imaginary
part in the denominator of the Feynman propagator at least near the poles of
unstable particles. In perturbation theory, this imaginary part
results from a resummation of self-energies.

To date there is no fully established treatment of unstable particles
within perturbation theory, although many solutions have been
proposed. The problem arises from the need to resum self-energies,
thus introducing a mixing of perturbative orders.
If done carelessly, this leads to violation of gauge
invariance and gauge independence. Thus, the naive modification of the
propagator to include a constant fixed width, the so-called
fixed-width scheme, violates Ward identities. When performing
precision calculations for Z production at LEP it turned out that
the renormalization of the Z-boson mass in the usual on-shell
renormalization scheme introduces a gauge dependence, as pointed out by
Stuart and Sirlin \cite{Stuart:1991,SirlinA:1991,SirlinB:1991}.

For inclusive observables that are dominated by the production of on-shell
unstable particles with a small width, finite-width effects can be neglected if
the required precision is small compared to the ratio of width and mass of the
unstable particles. This so-called narrow-width approximation is, however,
insufficient for many applications. 
A straight-forward gauge-invariant method for the inclusion of the finite width
is the factorization scheme introduced in \citere{BVZ:1992}, which consists in
the multiplication of the matrix elements with a global resonance factor.
However, for more complicated processes it becomes non-trivial to achieve a
precision beyond LO. 
The fermion-loop scheme \cite{Argyres:1995,Beenakker:1997} exploits the fact
that taking into account only closed fermion loops at the one-loop order allows
to perform a gauge-invariant and gauge-independent resummation. By construction
this method is restricted to leading-order predictions and to resonances that
decay exclusively into fermions. The idea of a gauge-invariant resummation can
be carried further by using the background-field method
\cite{Abbott:1980hw,Abbott:1983zw,Denner:1996wn} which allows to perform a Dyson
summation without violating Ward identities \cite{Denner:1996gb}. While the
resummed self-energies still depend on the quantum gauge parameter, this
dependence can be fixed by definition, \eg by using a specific gauge or the
prescription of the pinch technique \cite{PP:1995}.  In practice these methods
would require complete NNLO calculations to get NLO accuracy in the region of
the resonance.
The pole scheme proposed in \citeres{Stuart:1991xk,AO:1994} is based on the fact
that both the location of the pole and the residue of the propagator of an
unstable particle are gauge-independent. It allows to compute gauge-invariant
matrix elements to arbitrary orders via a Laurent expansion around the complex
pole. In practice this method gets quite involved in higher orders (see e.g.
\citere{Dittmaier:2014qza}), and usually only the leading terms in the Laurent
expansion are taken into account, called the leading-pole approximation.
Furthermore, effective field theory can be used to describe unstable particles.
In the method of \citeres{BBC:2000,FMZ:2013} non-local gauge-invariant effective
operators are introduced that allow the gauge-invariant resummation of
self-energies via appropriate choices of free parameters. In the
effective-field-theory approach of \citeres{Beneke:2003xh,Beneke:2004km} an
expansion in the coupling constant and in the distance from the pole is
performed simultaneously. This basically yields a field theoretically elegant
way to the pole approximation and can be easily combined with further expansions
(see \citere{BKS:2013} for a recent application).

The most straightforward method to describe unstable particles in
perturbation theory is the Complex-Mass Scheme (CMS)
\cite{Denner:2005,DDRW:1999,DD:2006}.  It is fully gauge-invariant,
valid everywhere in phase space, basically of the same complexity as a
calculation for stable particles and applicable to higher orders in
perturbation theory \cite{Actis:2006rc,Actis:2008uh}.  Finite widths
are introduced by analytically continuing the renormalized mass
parameters to appropriate complex values.  The introduction of complex
parameters immediately raises the question how unitarity is
implemented in this scheme.  Unitarity is not expected to be violated
because the bare Lagrangian is left untouched and only the
renormalization procedure is modified as compared to the standard
treatment.  Therefore any violation of unitarity should be beyond the
order of perturbation theory taken into account completely.  It has
been shown by Veltman \cite{Veltman:1963} within non-perturbative QFT
that unitarity is fulfilled in a theory with unstable particles
provided that the unstable particles are excluded from asymptotic
states.  Since the CMS provides a perturbative description of the full
theory it should not violate unitarity, if observables are correctly
computed in a valid perturbative regime.  Moreover, the CMS guarantees
exact gauge cancellations through gauge invariance order by order in
perturbation theory. Unitarity within the CMS has been touched upon in
\citere{Actis:2006rc}.  Unitarity in the CMS in a model with a heavy
vector boson interacting with a light fermion has been investigated at
the one-loop level in \citere{Bauer:2012gn}.

The aim of this paper is to study unitarity in scalar field theories in the CMS.
In Section 2 we shortly review unitarity and the Largest-Time Equation in the
case of stable particles and in Section 3 we summarise the Complex-Mass Scheme.
In Section 4 we investigate the realisation of unitarity in the Complex-Mass
Scheme by constructing and exploiting a suitable Largest-Time Equation for
unstable particles.


\section{Unitarity and Veltman's Largest-Time Equation for stable particles}


\subsection{Unitarity}
In the language of QFT unitarity means that the $S$ matrix is unitary, \ie
$\mathcal{S}^{\dagger} \mathcal{S} = \mathbbm{1}$. Separating the
non-interacting contributions from $\mathcal{S}$ via $\mathcal{S}=:
\mathbbm{1}+\imag \mathcal{T}$, one obtains the well-known relation
\begin{align}
        \mathcal{T}^{\dagger} \mathcal{T} = \imag
        \left(\mathcal{T}^{\dagger}-\mathcal{T}\right)
        \label{unitaritycond2}
\end{align}
for the transition matrix $\mathcal{T}$. A simple consequence of unitarity is
the optical theorem which states that the imaginary part of a forward scattering
amplitude $\mathcal{T}_{ii}$ is proportional to the total cross section:
\begin{align}
  \sigma_{\mathrm{tot}} = \text{flux factor} \times \Im{\mathcal{T}_{ii}}.
  \label{opticaltheorem}
\end{align}
The connection between \eqref{unitaritycond2} and the optical theorem
\eqref{opticaltheorem} is established when considering elements of the
transition matrix with definite initial and final states,
\begin{equation}
        \imag
        \left(\mathcal{T}_{if}^{*}-\mathcal{T}_{fi}\right) = \sum_{k}
        \mathcal{T}_{kf}^{*} \mathcal{T}_{ki},
        \specialnumber{a}\label{unitaritycond3}
\end{equation}
where the sum runs over all possible intermediate states $k$ and total
4-momentum conservation is implied. In scalar theories, where
$\mathcal{T}_{if} = \mathcal{T}_{fi}$ since the matrix elements do
not depend on the direction of the momenta, or in general for forward
scattering ($i = f$), the previous equation can be written as follows
\begin{equation}
  2 \Im{\mathcal{T}_{if}} = - 2 \Re{
    \raisebox{-18pt}{\includegraphics{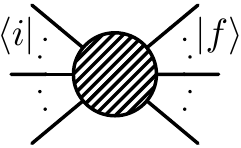}}
  } = \sum_k
  {\mathcal{T} \atop
\raisebox{+8pt}{\includegraphics{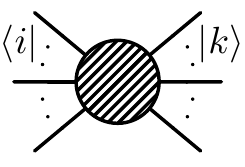}}}
{\mathcal{T}^* \atop
\raisebox{+8pt}{\includegraphics{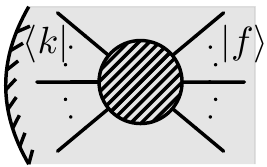}}} \;.
  \tag{\ref{unitaritycond3}} 
  \specialnumber{b}
\end{equation}
The so-called shadowed region is given by $\mathcal{T}^*$ while the normal
region is given by $\mathcal{T}$ and both transition amplitudes are connected by
on-shell states which is visualised as a cut (dark hatched line). We call the
equations \eqref{unitaritycond3} in the following the unitarity equation.
Unitarity is verified by computing the left-hand side of the unitarity equation
and comparing it to the right-hand side for all possible initial and final
states.  Direct computation of the left-hand side can be quite involved
especially beyond the one-loop level, but with the help of cutting rules, which
we introduce in the next section, the problem is solved theoretically and
practically.


\subsection{Veltman's Largest-Time Equation and unitarity}
\label{sec:LTE}
The LTE can be seen as the analogue to Cutkosky's cutting rules
\cite{CC:1960}, but is straightforward to derive and needs less
mathematical tools. The derivation of the LTE for stable particles can
be found, for instance, in Refs.  \cite{Veltman:1963,'tHooft:1973pz}
and it is based on a decomposition of the Feynman propagator in
space--time representation.  This decomposition is done, in the case
of stable particles, in positive- and negative-time parts in such a
way that positive (negative) time is connected to positive (negative)
energy flow and vice versa\footnote{The energy-flow direction is
  related to the sign of $p^0$, where $p^0$ is the zeroth component of
  the four-momentum. We prefer to speak about positive- and
  negative-time parts instead of positive- and negative-energy parts
  because of the generalisation to unstable particles introduced in
  \refse{se:decomposition}}. Let $\DeltaF(x-y)$ denote the Feynman
propagator of a scalar particle in space--time representation
\begin{align}
        \DeltaF(x-y) = \frac{1}{(2 \pi)^4} \int
        \mathrm{d}^4p \, \frac{e^{-\imag p(x-y)}}{p^2 -  m^2 + \imag
        \epsilon},
        \label{stableprop}
\end{align}
where $\imag \epsilon$ $(\epsilon>0)$ is an infinitesimal imaginary part
that ensures causality, then the decomposition is the following:\\
{\bf (Decomposition theorem)}: {\it There exist functions
  $\Delta^{\pm}$ with the properties}
\begin{align}
  \DeltaF\left(x_{i}-x_{j}\right)
  &=\theta\left(x_{i}^{0}-x_{j}^{0}\right)\Delta^{+}\left(x_{i}-x_{j}\right)
  +\theta\left(x_{j}^{0}-x_{i}^{0}\right)\Delta^{-}\left(x_{i}-x_{j}\right)
  \notag,\\
  \Delta^{\pm}\left(x_{i}-x_{j}\right) &= -\left(
  \Delta^{\mp}(x_{i}-x_{j})\right)^{*} = \Delta^{\mp}\left(x_{j}-x_{i}\right),
  \label{decompositiontheorem}
\end{align}
\noindent
{\it where $\Delta^{+}(x_{i}-x_{j})$ and $\Delta^{-}(x_{i}-x_{j})$
correspond to positive and negative energy flow, respectively.}

In Fourier space they take the simple form
\begin{align}
  \Delta^{\pm}\left(p,m^2\right) = \mp 2 \imag \pi \theta\left( \pm p_0\right)
  \delta\left(p^2 -m^2\right).
  \label{onshellpropagator}
\end{align}
Given such a decomposition, one can define extended Feynman rules:

\noindent
{\bf (The underline operation)}: {\it Given a Feynman diagram
  $\mathcal{F}$ defined by a set of vertices $\{x_i\}$ and
  corresponding couplings
  $\{g_i\}$, we define new diagrams where one or more of the
  space--time points $x_{i}$ can be underlined, \ie  $x_i\to\underline{x_{i}}.$
  This operation shall have the following consequences for propagators
  connecting the vertices in the original diagram:
  \begin{itemize}
  \item $\imag\Delta_{k i }=\imag\DeltaF(x_{k}-x_{i})$ {\it is unchanged if
      $x_{k},x_{i}$ are unchanged,}
  \item { \it$\imag\Delta_{k i }$ is transformed as $\imag\Delta_{k i }\to
        \imag\Delta_{k i
      }^{+}=\imag\Delta^{+}(x_{k}-x_{i})$ if $x_{k}\to \underline{x_{k}}$, but
  $x_{i}$ remains unchanged,}
  \item {\it $\imag\Delta_{k i }$ is transformed as
      $\imag\Delta_{k i }\to \imag\Delta_{k i
      }^{-}$ if $x_{i}\to \underline{x_{i}}$, but $x_{k}$ remains unchanged,}
  \item {\it $\imag\Delta_{k i }$ is transformed as $\imag \Delta_{k i }\to
    -\imag \Delta_{k i }^{*}$ if two connected space--time points $x_{k},x_{i}$
      are underlined,}
  \item {\it any underlined space--time point implies a factor $-1$
        for the corresponding vertex, \ie if $x_{k}\to
    \underline{x_{k}}$, then the corresponding coupling is replaced as
    $\imag g_{k} \to -\imag g_{k}$. }
  \end{itemize}
  }
  At the level of Feynman diagrams the underline operation is indicated by a
  circle $\bigcirc${} at the corresponding underlined space--time points. The
  rules stay the same for couplings with imaginary part, in particular, we
  stress that the coupling $g_i$ is not complex-conjugated for
  underlined $x_i$.
  
  As has been shown by Veltman in \citere{Veltman:1963} (see also
  \citere{'tHooft:1973pz}), the following equation can be derived from
  these rules:%
  \footnote{Equation \eqref{LTE} is actually a consequence of
    Veltman's LTE. For the sake of simplicity we use the term LTE for
    this equation in the following.}
  \\
{\bf(Largest Time Equation)}:{ \it Given a Feynman diagram $\mathcal{F}$ defined
  by a set of vertices $\{x_i\}$ and corresponding couplings $\{g_i\}$, if the
  Lagrangian is hermitian and all propagators fulfil the decomposition theorem,
  then the following equation holds}
  \begin{align}
  \sum_{\textrm{underlinings}}  \mathcal{F} (x_{1}, \ldots 
    \underline{x_i},\ldots, \underline{x_j},\ldots,x_N) =0,
    \label{LTE}
  \end{align}
  {\it where the sum runs over all possibilities of underlining
    elements $x_i$ (called LTE diagrams in the following).
In total there are $2^N$ contributions where $N$ is the number of vertices.}\\
\noindent
We note that the LTE holds both for truncated or non-truncated diagrams, and as
the prescription \eqref{LTE} is linear it does also hold for sums of diagrams
and for complete amplitudes $\mathcal{T}$.

The unitarity equation \eqref{unitaritycond3} is recovered by
extracting two contributions from the LTE, namely the one where none
of the vertices are underlined and the one where all of them are
underlined, \ie $\mathcal{T}(x_1, \ldots, x_i,$ $\ldots, x_N)$ and
$\mathcal{T}(\underline{x_1},\ldots,\underline{x_i},
\ldots,\underline{x_N})$.  These two contributions match $\imag
\mathcal{T}_{fi}$ and $-\imag \mathcal{T}_{if}^*$ and shifting them on
the other side of equation \eqref{LTE} we obtain the identity
visualised in step $\;\kreis{1}$ of \eqref{LTEplusKinematics}, where
the primed sum over underlinings stands for the sum over all possible
LTE amplitudes except for $\imag \mathcal{T}_{fi}$ and
$-\imag\mathcal{T}_{fi}^*$
\begin{align}
  -2\Re{
  \raisebox{-22.8pt}{\includegraphics{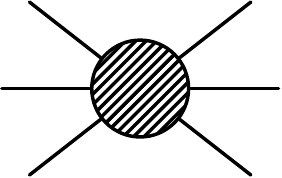}}}
  \underset{\;\;\kreis{1}}{=}
  \sum_{\mathrm{underlinings}}\nolimits^{\prime}
  \raisebox{-22.7pt}{\includegraphics{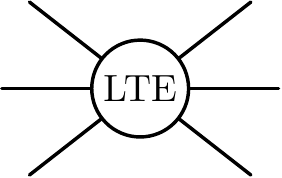}}
  \underset{\;\;\kreis{2}}{=}
  \sum_\mathrm{cuts}
  \raisebox{-24.5pt}{\includegraphics{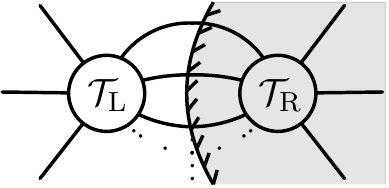}}
  \label{LTEplusKinematics}.
\end{align}
The right-hand side of \eqref{unitaritycond3}, \ie step $\;\kreis{2}$ of
\eqref{LTEplusKinematics}, is obtained by applying the kinematic
constraints, \ie the $\theta$ and $\delta$ functions, 
imposed by the explicit solutions $\Delta^{\pm}$
\eqref{onshellpropagator}. LTE amplitudes not satisfying these
constraints vanish and the remaining ones can be written in terms of
all possible ways of connecting amplitudes $\mathcal{T}_\mathrm{L}$
with complex conjugated amplitudes $\mathcal{T}_\mathrm{R}^*$ via cut
propagators (step $\;\kreis{2}$). The cut propagators $\Delta^\pm$ are
the solutions \eqref{onshellpropagator} which represent propagators
where one space--time point is not underlined while the other is.
Thus, each non-vanishing LTE amplitude has two well-defined regions, a
region with the usual Feynman rules which is always connected to the
incoming particles and a region with the ''complex-conjugated''
Feynman rules (underlined vertices) which is always connected to
outgoing particles.  Put in other words: Four-momentum conservation
and the given values of external four-momenta forbid certain
contributions to the LTE, and the contributions left are the ones
where the energy flows from incoming particles to outgoing particles
as it is required by the unitarity equation \eqref{unitaritycond3}.
The property of LTE amplitudes to fall apart into two separate
regions, thus justifying the representation of $\;\kreis{2}$, is
called {\em cut structure} in the following and reviewed in
\refse{se:kinematic}.

\noindent
{\bf(Cutkosky's cutting rules)}: The underline operation together with the LTE
are equivalent to Cutkosky's cutting rules, namely that the discontinuity of an
amplitude is obtained by replacing propagators in all possible ways by on-shell
propagators \eqref{onshellpropagator}, but constrained in such a way that the
energy flows from the initial to the final states. For more details, in
particular for the derivation of Cutkosky's cutting rules we refer to the
original reference \cite{CC:1960}.  In the following the terminologies
Cutkosky's rules, cutting rules and LTE with the usual on-shell cut propagator
\eqref{onshellpropagator} are used as synonyms.

\subsection{A decomposition for dressed propagators}
\label{se:bigpicture}

The applicability of the LTE is not restricted to amplitudes and can
be applied to Green's functions. A cut amplitude can be expressed in
terms of cut two-point functions which requires a decomposition,
similar to the case of the usual Feynman propagator, for two-point
functions. The decomposition can be achieved for dressed propagators
via the K\"all\'{e}n--Lehmann representation as has been shown by
Veltman in \citere{Veltman:1963}. Applying this idea to the unitarity
equation leads to a reinterpretation of the right-hand side of
\eqref{LTEplusKinematics} where simple cut propagators are replaced by
cut two-point functions
\begin{align}
  \sum_\mathrm{cuts}
  \raisebox{-25pt}{\includegraphics{cutamplitude.pdf}}
  &\underset{\;\;\kreis{3}}{=}
  \sum_\mathrm{cuts}
  \raisebox{-25pt}{\includegraphics{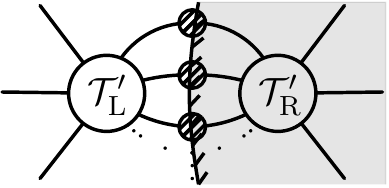}},
  \label{ltedressed}
\end{align}
where $\mathcal{T}_\mathrm{L,R}$ denote the subamplitudes on the left and
right-hand sides of the cut. 
The cut two-point function is given by
\begin{align}
   \raisebox{-10.5pt}{\includegraphics{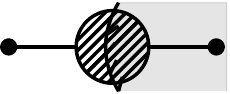}}
   =
   \raisebox{-10.5pt}{\includegraphics{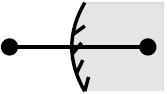}}
   +
   \raisebox{-10.5pt}{\includegraphics{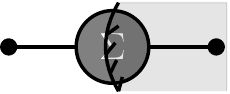}}
   +
   \raisebox{-10.5pt}{\includegraphics{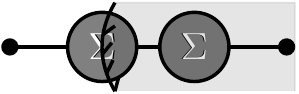}}
   +
   \raisebox{-10.5pt}{\includegraphics{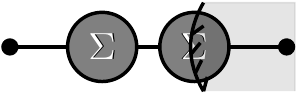}}
   +
   \ldots,
   \label{cuttwopointstable}
\end{align}
and the dotted lines in \eqref{ltedressed} indicate that we can have an
arbitrary number of cut propagators. The equality $\;\kreis{3}$ holds only for
the sum of all cut amplitudes. From the perturbative point of view the equality
follows by inserting the cut two-point function on the right-hand side of
\eqref{ltedressed} and by identifying $\mathcal{T}_\mathrm{L,R}$ with
$\mathcal{T}^\prime_\mathrm{L,R}$
supplemented by all non-cut parts of \eqref{cuttwopointstable}.


\section{The Complex-Mass Scheme}
When dealing with gauge theories it is crucial to guarantee gauge
invariance which is more involved when unstable particles are present.
As pointed out in the introduction, various methods have been
developed to describe unstable particles in perturbation theory, but
most are only valid near the resonance and lack validity in general
phase-space regions. In contrast, the CMS is valid in the full phase
space. Its underlying idea is an analytic continuation in
the masses of the unstable particles. Being analytic relations not
involving complex conjugation, the
Ward identities are not violated by such a modification. In practice,
the renormalized Lagrangian is rewritten by replacing any appearing
mass corresponding to an unstable particle with the complex one in
such a way that the bare Lagrangian is not changed. In a way the CMS
is just a renormalization scheme with complex renormalization
constants. 

We sketch the procedure: In the first step renormalized
parameters are introduced. Let $m_0$ denote the bare mass of an
unstable particle, then introduce
\begin{align}
  m_0^2  
  =:\mu^2+\delta \mu^2.
  \label{CMSCouplingRenormalization}
\end{align}
The complex mass $\mu^2$ is attributed to the propagator and resummed
while the counter term $\delta\mu^2$ is treated as a
vertex and not resummed. 

Thus, the LO propagator in the CMS reads
\begin{align}
  \DeltaF(x-y,\mu):=\frac{1}{(2 \pi)^4} \int\mathrm{d}^4p\;
  \frac{\mathrm{e}^{-\imag p (x-y)}}{p^2-\mu^2},
  \label{CMSPropagator}
\end{align}
or in momentum space
\begin{align}
  \DeltaF(p,\mu)=\frac{1}{p^2-\mu^2}.
  \label{CMSPropagator_mom}
\end{align}
The usual causality $\imag \epsilon$ prescription [see
\eqref{stableprop}] becomes irrelevant owing to the finite imaginary
part of $\mu^2=M^2-\imag \Gamma M$.

The procedure implies that the mass counter terms are
complex. Since the bare mass is real, the following
consistency equation holds
\begin{align}
  \Im{\mu^2}= -\Im{\delta \mu^2}.
  \label{CMSimagCT}
\end{align}
Couplings that are purely real in the conventional framework become
complex in the CMS if they are related to the masses, which is, for
instance, the case for the electroweak mixing angle in the
Glashow-Salam-Weinberg theory \cite{Denner:2005,DD:2006}.

We have to employ suited renormalization conditions in order to fix the finite
part of the parameters. Usually this is done in the on-shell scheme which is
distinguished by the fact that the renormalized parameters are equal to physical
observables. More concretely, one demands that the renormalized two-point
function of a stable particle near its mass $p^2=m^2$ is given by the Feynman
propagator \eqref{stableprop}. This condition does fix both the mass
renormalization and the field renormalization (see \eg \citere{Denner-renor}).
The on-shell scheme can be extended to the case of unstable particles, and the
appropriate renormalization conditions read \cite{Denner:2005,DD:2006}:
\begin{align}
  \left.\Sigma_\mathrm{R}\left(p^2\right)\right|_{p^2=\mu^2}
  =0,\qquad \left.\Sigma^\prime_\mathrm{R} 
  \left(p^2\right)\right|_{p^2=\mu^2} =0.
  \label{CMSRenormalizationCondition}
\end{align}
Here $\Sigma_\mathrm{R}$ denotes the renormalized self-energy of the unstable
particle and $\Sigma^\prime_\mathrm{R}$ is the corresponding renormalized
self-energy differentiated with respect to $p^2$. The renormalization conditions
\eqref{CMSRenormalizationCondition} together with the requirement that the bare
Lagrangian is real, yielding consistency equations like \eqref{CMSimagCT},
outline a gauge-invariant renormalization procedure. Apart from the validity of
Ward identities one must make sure that the renormalization conditions do not
introduce a gauge dependence. Given the fact that the complex pole is gauge
independent, the renormalization point and the renormalization condition
\eqref{CMSRenormalizationCondition} are gauge independent. Those proofs were
carried out by Stuart \cite{Stuart:1991}, Sirlin
\cite{SirlinA:1991,SirlinB:1991,Sirlin:1996}, Gambino and Grassi
\cite{Gambino:2000} and Grassi, Kniehl and Sirlin
\cite{Grassi:2001bz}.

Even though the renormalization conditions are similar to the ones in the
on-shell scheme the difference may be significant as it is the case in the SM
for the mass prediction of the W and Z bosons \cite{Bardin:1988xt}. In view of
gauge theories and physical observables, the complex pole is more than a
theoretical construct and should be seen as the analogue to the mass for stable
particles. For a discussion we refer to \citere{HVeltman:1994}.


\section{Unitarity in the CMS for scalar field theories}
\label{sec:unitaritycms}
In the CMS the Cutkosky rules are not valid in the sense that their application
does not yield the same result as one would get by direct computation of the
left-hand side of the unitarity equation \eqref{unitaritycond3}. For instance,
the Cutkosky rules require that the discontinuity of the tree-level $s$-channel
diagram vanishes for $s\neq m^2$,
\begin{align}
  -2\Ree{ 
  \mathop{\left[
  \vcenter{\hbox{\includegraphics{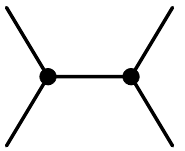}}}
  \right]}\limits_\text{\normalsize stable}}
  = 
  \begin{cases}
    0 & \text{if} \quad s \neq m^2\\
    \text{undefined} & \text{if} \quad s = m^2
  \end{cases}\; .
  \label{CutkoskyViolation}
\end{align}
Replacing the stable particle with an unstable one, the direct computation
yields
\begin{align}
  -2\Ree{ 
  \mathop{\left[
  \vcenter{\hbox{\includegraphics{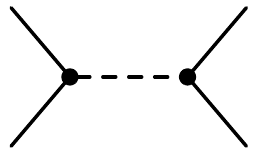}}}
  \right]}_\text{\normalsize unstable}} 
   = \frac{\Gamma M}{\left(s-M^2\right)^2 + \left(\Gamma M\right)^2} \neq 0 \quad
  \forall s .
\label{CutkoskyViolation2}
\end{align}
In view of the LTE the reason is that the preconditions are not fulfilled and we
cannot use the cutting rules for a propagator without having shown that there is
a valid decomposition \eqref{decompositiontheorem}.

As a consequence of the analytical continuation of the $S$ matrix to complex
masses algebraic relations are untouched, but operations where
complex-conjugation is involved are no longer preserved as it is the case for
the unitarity equation. The CMS guarantees gauge invariance, but it is no longer
clear how unitarity is implemented. Veltman has shown \cite{Veltman:1963} that
for a super-renormalizable theory the $S$ matrix in non-perturbative QFT is
unitary on the Hilbert space spanned by only stable particles,
\begin{align}
  \imag \left(\mathcal{T}_{if}^{*}-\mathcal{T}_{fi}\right)&= \sum_{
  \rb{k} \in \text{stable particles}} \mathcal{T}_{kf}^{*}
  \mathcal{T}_{ki}.
  \label{veltmanunitarity}
\end{align}
Starting from the K\"all\'{e}n--Lehmann representation for unstable
particles which, in contrast to stable particles, lacks a one-particle pole on
the real axis, he showed unitarity by deriving a LTE for dressed
propagators.

We apply this idea to the CMS in perturbative QFT, derive a corresponding LTE
and show that amplitudes $\mathcal{T}$ computed within the CMS at a perturbative
order $g^n$ for a fixed kinematic configuration are unitary up to higher orders,
\begin{align}
  \imag \left(\mathcal{T}_{if}^{*}-\mathcal{T}_{fi}\right)&= \sum_{
  \rb{k} \in \text{stable particles}} \mathcal{T}_{kf}^{*}
  \mathcal{T}_{ki}  + \mathcal{O}\left(g^{n+1}\right),
  \label{veltmanunitarity2}
\end{align}
where two sources of higher orders emerge. The first source are
kinematically suppressed terms in the LTE, which are effectively
suppressed by $\mathcal{O}(\Gamma)\sim\mathcal{O}(g^2)$, and the other
one are corrections due to the resummation of finite-width terms in
the CMS. Both turn out to result in non-relevant perturbative
corrections of order $\mathcal{O}\left(g^{n+1}\right)$.


\subsection{Sketch of the proof}
\label{se:sketch}

We restrict the discussion to a simple scalar toy model. It consists
of an unstable real scalar field $\phi$ with mass squared $\mu^2 = M^2
- \imag \Gamma M$ and a stable real scalar field $\chi$ with mass $m$
and the interaction 
\vspace{1em}
\begin{align}
  \mathcal{L}_\mathrm{I} = \frac{g}{2!} \phi \chi^2 \quad \Leftrightarrow \quad
  \raisebox{-19pt}{\includegraphics{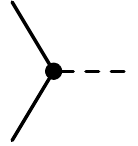}} = \imag g
  \;.
\end{align}

Owing to the required resummation of the width in the resonant
propagators there is no unique perturbative order for Feynman diagrams
or matrix elements in the CMS. The CMS propagator ${1}/(s-\mu^2)$
affects the perturbative order (it is of order $1/M\Gamma$ in the
resonance region but of order one otherwise), but has no Taylor
expansion near the resonance, \ie for $s \approx M^2$.  Since $\Gamma
M$ is of order $g^2M^2$, the change in perturbative order occurs near
the resonance.  The lack of a unique  perturbative order of
amplitudes  complicates the investigation of
perturbative unitarity in the CMS.  However, we can speak
of relative orders in the sense that if the left-hand side of the
unitarity equation has a phase-space-dependent order then also the
right-hand side does, and the difference in the orders is independent
of the phase-space region.

Our strategy concerning perturbative expansion is as follows: We rely on
perturbation theory in the CMS. We do never expand resonant propagators in the
amplitudes. We do, however, expand resonant cut propagators $\Delta^\pm$ (if not
stated otherwise, $\Delta$ denotes the propagator of the unstable particle in
the following). In our model, the width of the unstable particle is of order of
the coupling squared, \ie $\mathcal{O}(\Gamma)=\mathcal{O}(g^2)$. This allows to
define a consistent power counting in the region of a resonance and away from
it. While the absolute power counting depends on the phase-space region, the
relative power counting does not.

We here shortly sketch our derivation of Veltman's unitarity equation
\eqref{veltmanunitarity} in the CMS, to be elaborated in the following
subsections. It is done in four steps, similarly to the three steps
$\;\kreis{1}, \;\kreis{2}$ and $\; \kreis{3}$ of the equations
\eqref{LTEplusKinematics} and \eqref{ltedressed} described in
\refse{se:bigpicture}.  In \refse{se:decomposition} we construct a
decomposition of the CMS propagator of the form
\eqref{decompositiontheorem}.  Then, it follows immediately that we
can compute the left-hand side of \eqref{veltmanunitarity} via the LTE
[step $\; \kreis{1}$ in \eqref{ltecms}] as long as we consider
interactions with couplings that are either real or have a
corresponding complex-conjugated counterpart in the interaction part
of the Lagrangian\footnote{For real couplings that become complex in
the CMS, (\eg the mixing angles in the Standard Model) the LTE is
still valid because their complex-conjugated part is located in the
corresponding counter term which follows from the fact that the bare
coupling is real. Thus, any unitarity-violating terms from those
complexified couplings are trivially of higher order.},
\begin{align}
  -2\Re{
  \raisebox{-22.8pt}{\includegraphics{amplitude.pdf}}}
  &\underset{\;\;\kreis{1}}{=}
  \sum_{\mathrm{underlinings}}\nolimits^{\prime}
  \raisebox{-22.7pt}{\includegraphics{lteamplitude.pdf}}\notag\\
  &\underset{\;\;\kreis{2}}{=}
  \sum_\mathrm{cuts}
  \raisebox{-22.8pt}{\includegraphics{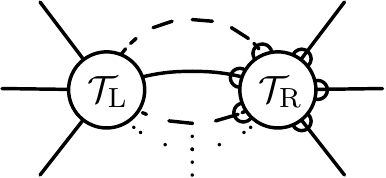}}
  +\mathcal{O}\left(g^{n+1}\right).
  \label{ltecms}
\end{align}
Otherwise the LTE is still valid, but the amplitude with all space--time points
underlined is not any more equal to the complex-conjugated amplitude, and the
left-hand side of the unitarity equation can no longer be related to
contributions to the LTE.

The preconditions for this identification are automatically fulfilled
for interaction vertices if the bare Lagrangian is real. The argument
fails for the imaginary mass counter term $\Im{\delta \mu^2}$ because
the corresponding counter part resides in the resummed propagator,
thus, we have a mismatch in the perturbative order and we have to take
care of this contribution differently.

In \refse{se:kinematic} we review kinematic arguments needed for
identifying cut contributions and we investigate the cut structure and changes
when unstable particles are present. 

In \refse{se:mass_ct} we explain how to compute the LTE amplitudes
when taking into account the imaginary mass counter term.
We show that the LTE amplitudes are obtained in the usual way except
for the cut contributions where each mass counter term insertion must
be rewritten appropriately.

In \refse{se:nonresonant} we show to all orders, using the
representation introduced in \refse{se:mass_ct}, that LTE amplitudes
within the CMS split into a normal region $\mathcal{T}$ and a
complex-conjugated region $\mathcal{T}^*$, up to terms of higher
perturbative order, \ie we show that step $\; \kreis{2}$ in
\eqref{ltecms} holds.  The circles attached to
$\mathcal{T}_\mathrm{R}$ in \eqref{ltecms} indicate that
$\mathcal{T}_\mathrm{R}$ is completely underlined (in the sense of
\refse{sec:LTE}). The plain (dashed) lines represent stable (unstable)
particles, and the dots indicate that we can have an arbitrary number
of cut propagators.  In order to identify higher orders we need to
expand in $\Gamma/M$. We do not perform the expansion for the whole
amplitude, but only for the cut propagators of the unstable particles
$\Delta^{\pm}$. We proof that LTE amplitudes containing kinematically
suppressed, or equivalently, non-resonant $\Delta^\pm$ do not
contribute to the unitarity equation in the considered order but are
always of higher, irrelevant order.

In \refse{se:resonant} we investigate resonant cut contributions.
Resonances in $\Delta^\pm$ play an essential role as they represent
kinematically allowed channels found on the right-hand side of the
unitarity equation. We point out that it is important to distinguish
the resonances appearing in the matrix elements
$\mathcal{T}_\mathrm{L}, \mathcal{T}_\mathrm{R}$ and resonant cut
contributions ($\Delta^\pm$) in view of unitarity. The former do
always appear in the same way on both sides of the unitarity equation,
thus enhancing both sides of the unitarity equation equally and we do
not need to consider them at all.  We stress that we only expand
unstable cut propagators, but never uncut propagators.  Expanding
unstable cut propagators at leading order in $\Gamma/M$, we find that
a resonant $\Delta^\pm$ of an unstable particle is just a $\delta$
function [see \eqref{stablelimit}] which seems to be in conflict with
the fact that unstable particles should not appear as asymptotic
states. However, we show that the $\delta$ function can be interpreted as
a cut self-energy of the unstable particle, thus being consistent with
Veltman's statement.

In order to proof unitarity in the CMS beyond one-loop we reformulate
the LTE in terms of nested LTE amplitudes of two-point functions in
\refse{sec:generalization} yielding the representation on the
right-hand side of the following equation:
\begin{align}
  \sum_\mathrm{cuts}
  \raisebox{-22.8pt}{\includegraphics{cutstructure.pdf}}
  \underset{\;\;\kreis{3}}{=}
  \sum_\mathrm{cuts}
  \raisebox{-22.5pt}{\includegraphics{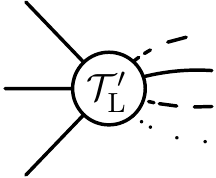}}
  \begin{matrix}
    \vspace{-0.8em}
    \raisebox{00pt}{\includegraphics[scale=0.8]{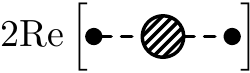}}\\
    \vspace{-0.8em}
    \raisebox{00pt}{\includegraphics[scale=0.8]{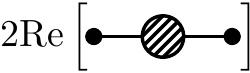}}\\
    \vspace{-1.4em}
    \raisebox{00pt}{\includegraphics[scale=0.8]{twopointunstable.pdf}}\\
    \quad\;\;\vdots
  \end{matrix}
  \raisebox{-25pt}{\includegraphics{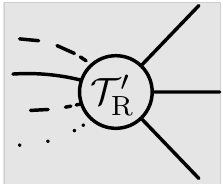}}
  +\mathcal{O}\left(g^{n+1}\right).
  \label{nestedtwopoint}
\end{align}
The leading-order result of our expansion (see \refse{se:resonant})
serves as the induction start for showing unitarity in general. From
\eqref{nestedtwopoint} unitarity is implied given that
\begin{align}
   \raisebox{-11.5pt}{\includegraphics{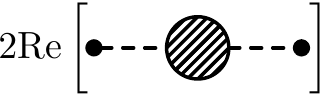}}
   \underset{\;\;\kreis{4}}{=}
   \raisebox{-10.7pt}{\includegraphics{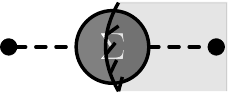}}
   +
   \raisebox{-10.8pt}{\includegraphics{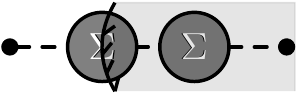}}
   +
   \raisebox{-10.8pt}{\includegraphics{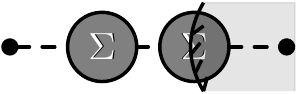}}
   \ldots +  \mathcal{O}\left(g^{n+1}\right)
   \label{cuttwopoint}
\end{align}
holds, and we recover the same expression as for stable particles
[\eqref{ltedressed} of \refse{se:bigpicture}]. In the final step we make use of
the fact that the CMS partially resums contributions and this partial
resummation appears in the mass counter term.  Rearranging the LTE in terms of
two-point functions, unitarity-violating terms cancel between self-energy terms
and mass counter terms (see \refse{sec:generalization}).


\subsection{Decomposition for the CMS propagator}
\label{se:decomposition}

We start with the construction of a propagator decomposition in the
case of the CMS. Since the decomposition is not unique, we list our
assumptions which led to it. Basically, one decomposes the
propagator into positive- and negative-time parts and adds something to
these parts which is zero for positive and negative times, respectively.
Working out this idea, one realises that the allowed transformations
have always, in Fourier space, the form of an advanced and retarded
propagator $\Delta_{\mathrm{A/R}}$. Hence, our approach consists of
defining meromorphic functions $\Delta_{\mathrm{A/R}}$ with similar
pole structure as in the case of stable particles, and among possible
restrictions Occam's Razor suggests
\begin{align}
        \DeltaF(p,\mu)-\DeltaA(p^{0},\vec p,M, \Gamma)\mathrel{\overset{!}{=}}
        \Delta^{+}(p^{0},\vec p,M, \Gamma),\nonumber\\
        \DeltaF(p,\mu)-\DeltaR(p^{0},\vec p,M, \Gamma)\mathrel{\overset{!}{=}}
        \Delta^{-}(p^{0},\vec p,M, \Gamma),\nonumber\\
        \int \mathrm{d}p^{0 }\;\Delta_{\mathrm{A/R}}(p^{0},\vec p,M, \Gamma)
        e^{\pm \imag p^{0} \abs{x^{0}}}= 0.
\end{align}
The function $\Delta_{\mathrm{A/R}}$ must be chosen such that
$\Delta^{\pm}$ fulfils the decomposition theorem
\eqref{decompositiontheorem}. The third equation is the condition that
the advanced/retarded propagator has only poles in the upper/lower
complex plane as it should be. Consequently, we have the same
situation as in the case of stable particles, namely $\theta(\pm
x^{0})\text{FT}\left[ \Delta_{\mathrm{A/R}}\right] (x)=0$, where FT
denotes the Fourier transformation. Furthermore, we demand that,
similar to the case of stable particles, the retarded propagator
turns into the advanced propagator by complex-conjugation and
vice versa, which is our last assumption
\begin{align}
  \DeltaA(p^{0},\vec p,M, \Gamma)&= \left(\DeltaR(p^{0},\vec p,M,
  \Gamma)\right)^{*}.
\end{align}
Given these restrictions one can easily derive the unique solutions for
$\Delta^\pm$. In Fourier space they read:
\begin{align}
  \Delta^\pm(p,\mu) = \imag \Im{\frac{1}{\hat p^0 \left(p^0 \mp \hat p^0
  \right) }} \quad \text{with} \quad \hat p^0 = \sqrt{ \vec p^2 + \mu^2}.
  \label{pseudocut}
\end{align}
As a first but very important result one verifies that in the limit $\Gamma\to
0^+$ our solutions turn into the stable ones, \ie 
\begin{align}
  \lim_{\Gamma \to 0^+} \Delta^{\pm}\left(p^2,\mu^2\right) = \mp 2 \pi \imag
  \theta\left(\pm p^0 \right) \delta \left( p^2 -M^2 \right).
  \label{stablelimit}
\end{align}
In view of consistency, this means that there is a smooth transition from
unstable to stable propagator as the mass $M^2$ tends below the kinematic limit
of instability. On the other hand, this result tells us that in a perturbative
expansion in $\Gamma$ the leading-order ''cut'' contribution is equal to the cut
contribution of a stable particle with the same mass. We note that
$\Delta^\pm(p,\mu)$ for finite $\Gamma$ does neither involve a $\delta$ nor a
$\theta$ function. Thus, energy can flow in both directions and the realisation
of causality is more involved for unstable particles.

Apparently, two problems appear:
\begin{itemize}
\item Given an $S$ matrix, an
  expansion in small $\Gamma$ can often be performed only in a
  distributional sense, even though perturbation theory
  predicts $\mathcal{O}(\Gamma) = g^2$, where $g$ is the coupling
  constant. For instance, for the $s$-channel production of an
  unstable particle the width is crucial for the finiteness of the
  result. The question is when we are allowed to do a naive expansion
  or when is it actually necessary since we only want to verify
  the unitarity equation \eqref{unitaritycond3}.
\item At first sight the fact that the cutting rules for the CMS
  propagator for $\Gamma \to 0^+$ coincide with the ones for stable
  particles might interfere with Veltman's result, namely that only
  stable particles appear as asymptotic states in the unitarity
  equation \eqref{veltmanunitarity}.
\end{itemize}
Both points do not pose any problems, as we show in the upcoming sections.


\subsection{Cutting rules for unstable particles}
\subsubsection{Kinematic restrictions}
\label{se:kinematic}
The cutting rules are a special case of the LTE relations where many terms in
\eqref{LTE} do not contribute because the $S$ matrix underlies physical
constraints such as positive energy flow and real masses. These constraints
reappear in the LTE amplitudes in form of $\delta$ and $\theta$ functions.

The situation is similar for stable and unstable particles, and we consider
stable particles first. In our convention the incoming particles are on the left
and the outgoing ones on the right. As an example consider the following diagram
in a scalar $\phi^3$ theory:
\begin{align}
  \mathcal{F}\left(p_1, p_2, p_3, p_4 \right) = 
  \raisebox{-28.5pt}{\includegraphics{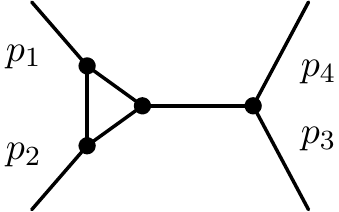}} \;\;.
  \label{ex2}
\end{align}
If we sum over all possible underlinings the result equals zero except
for one contribution which is immediately recognised as a cut
\begin{align}
  -2 \Re{\mathcal{F}} = 
  \raisebox{-29pt}{\includegraphics{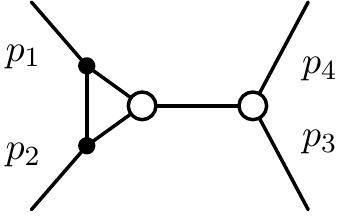}}
  :=
  \raisebox{-29pt}{\includegraphics{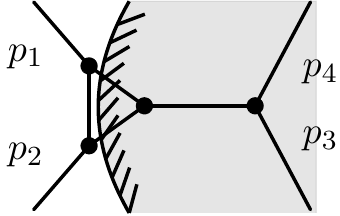}} \;\;.
  \label{ex5}
\end{align}
The example shows that the non-vanishing contributions of the LTE for
stable particles split into two separate regions where the normal part
($\mathcal{T}$) is given by the black dots, while the
complex-conjugated part ($\mathcal{T}^*$) is given by the white
circles. In the following we call this property the {\it cut
  structure}. This means, in particular, that for stable particles the
LTE and simple kinematic arguments lead to the unitarity equation
\eqref{unitaritycond3}. Examples of vanishing LTE terms are
\begin{align}
  {\theta\left(-p_0 \right) \atop 
  \raisebox{10pt}{\includegraphics{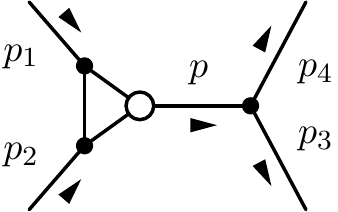}}} = 0 \;, \qquad
  {\delta\left(p^2 - m^2 \right) \atop 
  \raisebox{10pt}{\includegraphics{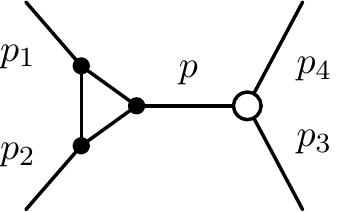}}} = 0 \;.
  \label{ex3}
\end{align}
In the first term of \eqref{ex3} the cut structure is violated, \ie
there are no two well-defined regions, and as a consequence at some
vertex in the amplitude the required energy-flow direction is opposed
to the physical energy flow which is from the left to the right, and
no energy can be transferred to the final state. The second amplitude
vanishes because the cutting rules require the intermediate particle
to be on-shell which is impossible for stable particles since $m^2 <
p^2 = (p_1+p_2)^2$.

When CMS propagators are involved one would like to have, in particular, the cut
structure of non-vanishing LTE terms, but this is a priori not given. For
instance, the first term in \eqref{ex3} does not vanish when the cut propagator
is replaced by the corresponding CMS propagator. These cut-structure-violating
contributions come from the fact that for the CMS propagator $\Delta^\pm(p,\mu)$
\eqref{pseudocut} there is neither a $\theta \left(\pm p^0\right)$ nor a
$\delta\left(p^2-M^2\right)$ but smoothed functions instead. The smoothing does
no longer enforce the same strict kinematic constraints as for stable particles.
Nevertheless, these contributions are suppressed by at least a factor
$\Gamma/M\sim g^2$, and one obtains the same behaviour for unstable particles in
a perturbative sense, meaning that those LTE terms violating the cut structure
are always of higher order in the coupling constant.  While the first LTE
diagram in \eqref{ex3} is also perturbatively suppressed for $p^2\sim M^2$, the
second one does not violate the cut structure and, in fact, in the case of
unstable particles its contribution is relevant.  Such contributions are
discussed in \refses{se:resonant} and \ref{sec:generalization}.

Yet, the argument is incomplete since suppressed terms can become relevant as
one takes into account higher perturbative orders, \ie they can be of the same
order as higher quantum corrections. The next chapter is devoted to include the
imaginary mass counter term in the LTE. We discuss how to simplify LTE relations
and we show that the imaginary mass counter term is responsible for the fact
that contributions being negligible at a certain perturbative order, stay
negligible even if the calculation is extended to higher orders (see
\refse{se:nonresonant} for non-resonant and \refses{se:resonant} and
\ref{sec:generalization} for resonant propagators).


\subsubsection{Including the imaginary mass counter term}
\label{se:mass_ct}

A proper description of unstable particles requires the resummation of
self-energy contributions resulting in a non-zero imaginary part in
the LO propagator. On the other hand, gauge invariance requires that
the imaginary counter part of the complex mass enters the Feynman
rules. It is not possible \cite{tHooft:2005} to include such a
coupling in the LTE relations. However, this is not necessary as we
discuss in the following. Consider the insertion of a $\imag (-\imag
\Gamma M)$ coupling between two CMS propagators in momentum space,
\begin{align}
  \left( \imag \Delta \right) \left( - \imag^2 \Gamma M \right)\left( \imag
  \Delta \right)
  = \raisebox{-1pt}{\includegraphics{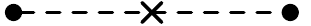}} \;.
  \label{unstableprop2}
\end{align}
This insertion can always be reduced to the usual propagator via simple
differentiation with respect to $\Gamma M$
\begin{align}
  \left(\imag \Delta\right) \left(-\imag^2 \Gamma M\right) \left(\imag
  \Delta\right) = -\Gamma M \frac{\partial}{\partial \Gamma M} \imag \Delta,
  \label{undoresum}
\end{align}
and, as it becomes clear below, it is important that $\Gamma M$ is real which is
true by construction. 

For an arbitrary amplitude $\mathcal{T}$, the left-hand side of the unitarity
equation \eqref{unitaritycond3} is computed as follows: Every insertion of
$\imag(-\imag \Gamma M)$ can be generated by differentiating specific CMS
propagators according to \eqref{undoresum}. Consequently, any
amplitude with $(-\imag \Gamma M)$ insertion in the
CMS can be generated from the diagrams in an amplitude where $\imag(-\imag
\Gamma M)$ is missing. Consider the diagrams $\mathcal{F}^\tau$ obtained from
$\mathcal{T}$ by setting the imaginary mass counter term $\Gamma M$ to zero, but
keeping the resummed counter part in the propagator
\begin{align}
  \left. \mathcal{T} \right|_{\Gamma M = 0} = \sum_\tau \mathcal{F}^\tau.
\end{align}
We denote the set of propagators which are linked to a mass counter term by
$\Omega :={} \{ \Omega_{\tau,i}\}$ where $i$ identifies the
propagator and $\tau$ the diagram in $\mathcal{T}$ which contains the
propagator $i$. Multiple insertions of $\imag(-\imag \Gamma M)$ are collected
and the number of successive insertions is denoted by $n_{\tau}^i$.  We
transform each propagator $i$ in $\tau$ via $\Phi_\Omega:\Gamma M \to \Gamma M
+\omega_{\tau}^i$ in its denominator, and the amplitude $\mathcal{T}$ is then
generated via
\begin{align}
  \mathcal{T} = \left.\sum_{\tau \in \Omega} \prod_{i \in \tau}
  \frac{1}{n_{\tau}^i!}\left(-\Gamma M
  \frac{\partial}{\partial\omega_{\tau}^i}\right)^{n_{\tau}^i}
  \mathcal{F}^{\tau}_{\Phi_\Omega} \right|_{\omega_{\tau}^i = 0},
  \label{labelingF}
\end{align}
where $\mathcal{F}^{\tau}_{\Phi_\Omega}$ is the diagram $\mathcal{F}^{\tau}$
with the propagators transformed according to $\Phi_\Omega$.

By construction $\mathcal{F}^{\tau}_{\Phi_\Omega}$ is free of
imaginary mass counter terms and in the case of a scalar theory, as we
consider it here, we can simply commute differentiation with taking
the real part. Thus, we can directly apply the LTE on
$\mathcal{F}^{\tau}_{\Phi_\Omega}$ and after performing the (real)
differentiation we obtain the real part of $\mathcal{T}$ via
\eqref{labelingF}. This representation implies that the cutting rules
in $\mathcal{T}, \mathcal{T^*}$ stay the same even if the imaginary
mass counter term is included. To see this consider the case when we
have a CMS propagator and at least one insertion of $\imag(-\imag
\Gamma M)$ in an amplitude $\mathcal{T}$
\begin{align}
  \mathcal{T} = \int \tilde{\mathcal{T}} 
  \;\imag g\;\imag \Delta \left(p^2, M^2 - \imag \Gamma M\right) 
  \; \imag(-\imag \Gamma M)
  \; \imag \Delta \left(p^2, M^2 - \imag \Gamma M\right)\;\imag g,
\end{align}
where $\tilde{\mathcal{T}}$ denotes the amplitude $\mathcal{T}$ with the
two-point function with the imaginary mass counter term insertion omitted and
the two couplings $(\imag g)^2$ connecting $\tilde{\mathcal{T}}$ and the
two-point function between those couplings removed.  The $\int$ indicates that
the propagator's momentum $p^2$ may be integrated.  Rewriting the insertion one
is left with
\begin{align}
  \mathcal{T} = -\Gamma M \frac{\partial}{\partial \omega} 
  \underbrace{\int\tilde{\mathcal{T}} 
  \;\imag g\;\imag \Delta \left(p^2, M^2 - \imag \left(\Gamma M +
  \omega\right)\right)\;\imag g}_{=
  \mathcal{T}^{\tau}_{\Phi_\Omega}}\biggl|_{\omega = 0}.
\end{align}

Applying the LTE, the propagator $\imag \Delta$ either stays the same ($\in
\mathcal{T}$), transforms into $-\imag \Delta^*$ ($\in \mathcal{T}^*$), or
belongs to the cutting region
\begin{align}\label{eq:propagator_LTE}
0
= -\Gamma M \frac{\partial}{\partial \omega}
  \int  
  \biggl(\sum_{\mathrm{underlinings}}
  \tilde{\mathcal{T}}\biggr)
  \;\left(\imag g\right)^2\bigl[&\imag \Delta \left(p^2, M^2 - \imag
  \left(\Gamma M + \omega\right)\right)\nonumber\\
  &{}+(-1)^2\times(-\imag)\times\Delta^* \left(p^2, M^2 - \imag \left(\Gamma M +
  \omega\right)\right)\nonumber\\
  &{}+(-1)\times\imag \Delta^+ \left(p^2, M^2 - \imag \left(\Gamma M +
  \omega\right)\right)\nonumber\\
  &{}+(-1)\times\imag \Delta^- \left(p^2, M^2 - \imag \left(\Gamma M +
  \omega\right)\right)
\left. \bigr]\right|_{\omega = 0}.
\end{align}
While the first term in \refeq{eq:propagator_LTE} corresponds to the
propagator left of the cut and the second term to the one right of the
cut, the third and fourth terms represent cut propagators 
with dominantly positive or negative energy flow.
The sum over the underlinings of $\tilde{\mathcal{T}}$ represents the LTE
equation of the truncated amplitude\footnote{The amplitude $\tilde{\mathcal{T}}$
is missing two vertices compared to $\mathcal{T}$, not counting the imaginary
mass counter term.
Therefore, the total number of LTE amplitudes
corresponding to $\tilde{\mathcal{T}}$ is $2^{n-2}$ where $n$ is the number of
vertices in $\mathcal{T}$.  Multiplying with the four underlinings of the
propagator we recover the $2^n$ LTE amplitudes of $\mathcal{T}$. }
$\tilde{\mathcal{T}}$.  If the propagator is in the normal region we recover the
result \eqref{undoresum} after applying the differentiation
\begin{align}
  \left.-\Gamma M \frac{\partial}{\partial \omega}
  \imag \Delta\left(p^2, M^2 - \imag \left(\Gamma M +
  \omega\right)\right)\right|_{\omega = 0} =
  \imag \Delta\left(p^2,\mu\right)
  \imag\left(-\imag \Gamma M\right)
  \imag \Delta\left(p^2,\mu\right).
\end{align}
For the propagator in the complex-conjugated region $- \imag \Delta^*(p,\mu)$ we
can work out the signs leading to
\begin{align}
  \left.-\Gamma M \frac{\partial}{\partial \omega}
  \left(-\imag \Delta^*\left(p^2, M^2 - \imag \left(\Gamma M +
  \omega\right)\right)\right)\right|_{\omega = 0} =
  \left(-\imag \Delta^*\left(p^2,\mu\right)\right)
  \imag\left(-\imag \Gamma M\right)
  \left(-\imag \Delta^*\left(p^2,\mu\right)\right).
\end{align}
Consequently, we obtain cutting rules in the regions $\mathcal{T},\mathcal{T}^*$
for $\imag (-\imag \Gamma M)$ which coincide with the usual Feynman rules,
namely that $\imag (-\imag \Gamma M)$ is treated purely real.  

Remember that we cannot deal with {\it arbitrary} complex couplings in the LTE
unless we give up the relation between LTE and unitarity. Imaginary couplings do
not turn into their complex conjugates according to the last underlining rule
(see \refse{sec:LTE}) which is necessary for identifying the left-hand side of
the unitarity equation, but we have shown that $\imag (-\imag \Gamma M)$
transforms correctly by other means.  The result is true for more than just one
insertion which can be shown by working out the signs for multiple
differentiation. The case that the propagator of the unstable particle is on the
cut is discussed in the following sections. 


\subsubsection{Non-resonant contributions of unstable particle propagators}
\label{se:nonresonant}

We come back to the question whether contributions, which in the case of stable
particles vanish because of kinematic constraints, can actually contribute in
the case of unstable particles.  In this subsection we show that the
$\imag(-\imag \Gamma M)$ insertions make sure that contributions that vanish for
stable particles never become relevant for unstable ones (that are not
resonant).

Consider the amplitude
\begin{align}
  \imag \mathcal{T} =
  \underset{\kreis{1}}
  {\raisebox{-19pt}{\includegraphics{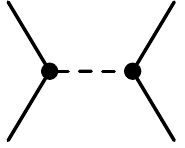}}}
  +
  \underset{\kreis{2}}
  {\raisebox{-19pt}{\includegraphics{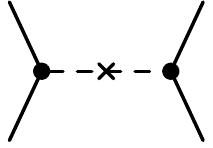}}}
  +
  \underset{\kreis{3}}
  {\raisebox{-19pt}{\includegraphics{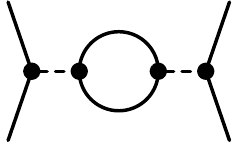}}}
  \label{SChannelOneLoop}
\end{align}
and assume $\left|p^2 -M^2\right| \gg M\Gamma$ (off resonance), then the
order of accuracy of the amplitude is $\mathcal{O}(\mathcal{M}) =
g^4$. Computing the unitarity equation \eqref{unitaritycond3}, the
leading contribution to the left-hand side results from \;$\kreis{3}$
\begin{align}
  -2 \Re{
    \raisebox{-20pt}{\includegraphics{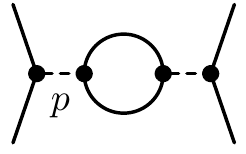}}
    } =
    \raisebox{-24.5pt}{\includegraphics{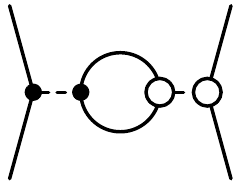}}
    \left( 1 + \mathcal{O}\left(g^2 \right) \right)=
    \raisebox{-24.5pt}{\includegraphics{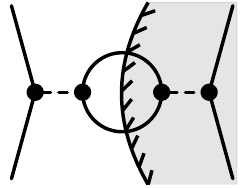}}
    \left( 1 + \mathcal{O}\left(g^2 \right) \right) \;.
\end{align}
The higher-order contributions $\mathcal{O}\left(g^2\right)$ are of the type of
\eqref{ex3}, \ie they either violate the cut structure or are further suppressed
[owing to $p^2 \neq M^2$ and \refeq{eq:Deltapm_expansion}], but they have the
topology of \;$\kreis{3}$.

As we have demonstrated in \refse{se:mass_ct}, we can take into account the
imaginary mass counter term via differentiation, and the left-hand side of the
unitarity equation \eqref{unitaritycond3} for \;$\kreis{1}+\;\kreis{2}$ reads
\begin{align}
  \left(1 - \Gamma M \frac{\partial}{\partial \Gamma M} \right) \left[
  \raisebox{-19.5pt}{\includegraphics{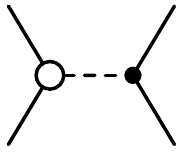}}
  +
  \raisebox{-19.5pt}{\includegraphics{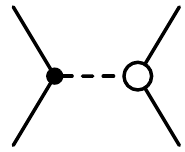}}
  \right].
  \label{SChannelOneLoop1and2}
\end{align}
The first term can never become resonant because it violates the cut
structure and the second term is non-resonant as long as the
amplitude itself is non-resonant which is true by our assumption $s \not\approx
M^2$. Deriving the leading behaviour of $\Delta^\pm(p,\mu)$ for small
$\Gamma/M$ for $p_0 \neq \pm \sqrt{\vec p^2 +M^2}$ we obtain
\begin{align} 
  \Delta^\pm(p,\mu) 
  &= \pm \imag \frac{\left(-p_0 \pm 2 \sqrt{\vec p^2 +
  M^2}\right)}{2 \sqrt{\vec p^2 + M^2}^3 \left( p_0 \mp \sqrt{\vec p^2 +
  M^2}\right)^2} \Gamma M +\mathcal{O} \left(\left(\frac{\Gamma}{M}\right)^3
  \right) \nonumber\\
  &= \pm \Gamma M f\left(\Gamma M \right),
\label{eq:Deltapm_expansion}
\end{align}
where $f$ is a smooth function with Taylor expansion in $\Gamma M$. The explicit
factor in $\Gamma M$ indicates the suppression and after carrying out the
differentiation in \eqref{SChannelOneLoop1and2} the leading term of order
$\Gamma M$ is eliminated and the resulting order is $g^2 \times
\mathcal{O}(\Gamma^2) = g^6$.  Since $g^4$ is the current accuracy of the
amplitude \eqref{SChannelOneLoop} the contributions from
\;$\kreis{1}+\;\kreis{2}$ are negligible, which is no coincidence.

The argument is easily extended to arbitrary high order. Consider an amplitude
up to the order of $g^n$, where $n$ is arbitrary. We compute the LTE according
to \eqref{labelingF} and assume we have a term $\mathcal{U} \in
\mathcal{F}^{\tau}$ of the order of $g^m$ and $m \le n$ either violating the cut
structure or having at least one non-resonant $\Delta^\pm(p,\mu)$. The order of
$\mathcal{U}$ is bounded from below as follows
\begin{align}
  \mathcal{O}(\mathcal{U}) \ge g^m \left. \sum_{k=0}^{\frac{n-m}{2}}
  \frac{1}{k!} \left(-\xi \frac{\partial}{\partial \Gamma M} \right)^k \Gamma M
  f (\Gamma M)\right|_{\xi =\Gamma M} 
  = \mathcal{O}\left(g^{m}\Gamma^{\frac{n-m}{2}+1} \right)
  = \mathcal{O}\left(g^{n+2} \right),
  \label{errorbound}
\end{align}
where the equality $(=)$ occurs solely for one non-resonant $\Delta^\pm$ in
$\mathcal{U}$. Multiple insertions of $\imag(-\imag \Gamma M)$ result in a
systematic elimination of orders as can be easily seen realising that the
differential operator in \eqref{labelingF} is nothing but the Fourier
representation of the translation operator
\begin{align}
  \left(\mathrm{e}^{-\xi \frac{\partial}{\partial \Gamma
  M}}\right)_{\frac{n-m}{2}} := \sum_{k=0}^{\frac{n-m}{2}}
  \frac{1}{k!}\left(-\xi \frac{\partial}{\partial \Gamma M}\right)^{k},
\end{align}
where the series of the exponential function is terminated at the order $(\Gamma
M)^\frac{n-m}{2}$. On the other hand, the translation operator acts as follows
on a function $P$: $\left.\mathrm{e}^{-\epsilon_{0} \frac{\partial}{\partial
\epsilon}} P(\epsilon)\right|_{\epsilon_{0}=\epsilon} = P(0)$. Thus, we obtain
the same result with possible deviations starting at the order $(\Gamma
M)^{\frac{n-m}{2}+1 } = \mathcal{O}\left(g^{n-m+2} \right)$ which shows the
result \eqref{errorbound}. 

Loosely speaking, the non-resonant propagators are expanded in
$\Gamma$ and thereafter the resummed and non-resummed terms explicitly
cancel.  As finite-width terms in the complex-mass counter term
$\delta\mu^2$ result only from a reparametrisation of the theory,
resummed and non-resummed terms have to compensate each other in each
fully calculated order.

The results so far can be summarised as follows:
\begin{itemize}
  \item There exists a decomposition for the CMS propagator \eqref{pseudocut}
    satisfying the decomposition theorem \eqref{decompositiontheorem}, thus,
    allowing to derive a LTE [step $\;\kreis{1}$ in \eqref{ltecms}].
  \item The LTE does not allow to include the imaginary mass counter term
    directly, but it can be introduced via \eqref{labelingF}.
  \item We have shown that cut-structure-violating terms as well as
    all cuts of non-resonant propagators can always be neglected no
    matter at which order in the coupling constant the violation takes
    place. Thus, only correctly cut LTE amplitudes have to be taken
    into account which is required by unitarity [step $\;\kreis{2}$ in
    \eqref{ltecms}].
\end{itemize}

Further, from the stated results it follows immediately that unitarity
is fulfilled automatically if there are no resonant
$\Delta^\pm(p,\mu)$. The missing piece which has yet to be
investigated is when $\Delta^\pm(p,\mu)$ becomes resonant. This
happens when internal momenta are integrated out, or usually when
certain phase-space integrations are carried out.


\subsubsection{Resonant contributions of unstable particle propagators at
one-loop order}
\label{se:resonant}

In this section we discuss resonant $\Delta^\pm(p, \mu)$ at leading order in
$\Gamma/M$. Those terms are no longer negligible in the LTE as for instance the
second term of \eqref{ex3} for an unstable $s$-channel particle and $s \approx
M^2$.

Naively interpreting the unitarity equation \eqref{unitaritycond3}
would lead us to the conclusion that only the sums of all diagrams on
both sides of the unitarity equation coincide, though, in the case of
stable particles diagrams can be separated according to their topology
and perturbative order.  Perturbative unitarity then follows from the
fact that the coupling can be chosen arbitrarily meaning that we can,
in principle, distinguish between orders by varying the coupling. This
argument can not be directly transferred when the theory is
renormalized according to the CMS.  The distinction of perturbative
orders does no longer work because of resummation, and we actually
have to consider sums of diagrams, but the occurrence of non-trivial
relations between topologically different Feynman diagrams can be
excluded at least in scalar theories.  As discussed in
\refse{se:sketch}, we can consider relative orders in the sense that
if the left-hand side of the unitarity equation has a
phase-space-dependent order then also the right-hand side does, and
the difference of orders is independent of phase space.  Our strategy
is therefore to identify the diagrams not only by their topology, but
also by the perturbative order in certain phase-space regions where it
is well-defined, \eg for $\left|s - M^2 \right| \sim \Gamma M$ or
$\left|s-M^2\right|\gg \Gamma M$.

At this point we recall that besides the loop expansion we only expand cut
propagators $\Delta^\pm$ connecting the regions $\mathcal{T},\mathcal{T}^*$ in
$\Gamma/M$, but do never expand the propagators  $\Delta$ within
$\mathcal{T}$ or $\mathcal{T}^*$.

We start again with the example of the $s$-channel production of an unstable
particle. For the resonant case one must perform a Laurent expansion to capture
the leading behaviour,
\begin{align}
  \left.\Delta^\pm(p,\mu)\right|_{p_0 = \pm \sqrt{\vec p^2 +M^2}} = \mp
  \imag\frac{2 }{\Gamma M} + \mathcal{O}\left( \frac{\Gamma}{M}\right).
  \label{cutpropLaurent}
\end{align}
The LTE at LO reads for $p^2 = M^2$ \nopagebreak
\begin{align}
  -2 \Re{
   \raisebox{-20pt}{\includegraphics{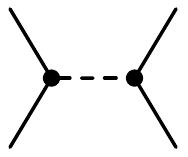}}
  }
  =
  \raisebox{-20pt}{\includegraphics{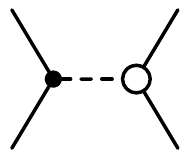}}
  \left( 1 + \mathcal{O}\left(g^2\right) \right)
  =
  \raisebox{-20pt}{\includegraphics{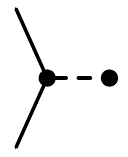}}
  \left( 2 \Gamma M \right)
  \raisebox{-20pt}{\includegraphics{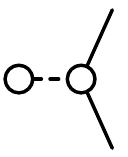}}
  \left( 1 + \mathcal{O}\left(g^2\right) \right) \;,
  \label{SChannelTree}
\end{align}
where in the last step we made use of \eqref{cutpropLaurent} and the identity
\begin{equation}
  \frac{1}{\Gamma M} = \left.\Delta(p,\mu) \Gamma M
  \Delta^*(p,\mu)\right|_{p^2=M^2}.
  \label{eq:identity}
\end{equation}
The LTE does not know about the diagrammatic significance of $\Gamma M$ which we
have to determine and plug in by hand, and, as discussed before, $\Gamma M$ must
be computed at least by the one-loop renormalization conditions
\eqref{CMSRenormalizationCondition}. We denote $\Sigma_\mathrm{R}^1$ the
renormalized (according to the CMS) one-loop self-energy of the unstable
particle and $\Sigma^1$ the corresponding unrenormalised one. The self-energies
are related to the counter term $\delta\mu^2$ by the renormalization conditions
\eqref{CMSRenormalizationCondition}, and the consistency equation
\eqref{CMSimagCT} links $\delta\mu^2$ and $\Gamma M$, so we obtain a relation
between $\Gamma M$ and $\Sigma^1$
\begin{equation}
  \Gamma M = \Im{\delta \mu^2}=  -\Re{ \left.\imag \Sigma^{1}(p^2)
  \right|_{p^2 = \mu^2}}.
  \label{GMrelation}
\end{equation}
In the next step we assume that the analytic continuation of the self-energy
behaves well enough at $p^2 \approx M^2$, \ie we suppose that
\begin{equation}
  \left. \frac{\imag \Sigma_\mathrm{R}(p^2)}{\Gamma M} \right|_{p^2 \approx M^2}
  = \mathcal{O}\left(g^2\right),
  \label{sigmacondition}
\end{equation}
which can be obtained formally by performing an expansion in $p^2$, but
sometimes a Taylor expansion is not possible as it is the case when infrared
singularities appear. Then, one usually has logarithmic corrections, but they do
not bother us as long as the limit $g \to 0$ exists. In the next step we make
use of the assumption \eqref{sigmacondition} and find for $\Gamma M$:
\begin{align}
  \mathcal{O}\left( g^4 \right)&= \Re{\left.\imag
  \Sigma_{\mathrm{R}}\right|_{p^2 \approx M^2}}= \Re{ \left. \imag \Sigma
  \right|_{p^2 \approx M^2}} + \Gamma M \nonumber\\
  \Rightarrow \Gamma M &= - \Re{\left.  \imag \Sigma \right|_{p^2 \approx M^2}}
  +\mathcal{O} \left( g^4 \right) =- \Re{\left. \imag \Sigma\right|_{p^2 \approx
  M^2}} \left( 1+\mathcal{O}\left( g^2 \right) \right). 
\end{align}
This equation expresses what is known from the on-shell scheme, \ie the width is
the cut through loops and can be interpreted as the decay width. At one-loop
order the widths in the CMS and the on-shell scheme coincide, but this is no
longer true at higher orders and we will not be able to argue this way in the
general case.

Nevertheless, let us make use of this result to demonstrate
unitarity at one-loop order. At this order we can directly apply the cutting
rules to $\Sigma^1$ since there are no intermediate unstable particles (in our
model)
\begin{align}
  -2 \Re{\imag \Sigma^1} =
  \raisebox{-16.7pt}{\includegraphics{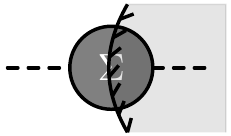}}
  =
  \raisebox{-16.7pt}{\includegraphics{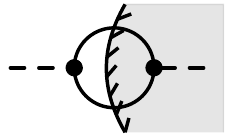}} \; .
  \label{LTECutoneloopSelfEnergy}
\end{align}
This result together with \eqref{SChannelTree} and \eqref{GMrelation} yields
exactly what is required by unitarity, namely
\begin{align}
  -2 \Re{
    \raisebox{-19.5pt}{\includegraphics{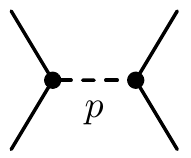}}
  }
  =
  \left( \raisebox{-24.8pt}{\includegraphics{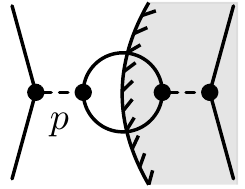}}
  \right) \left(1 + \mathcal{O} \left(g^2\right) \right), \quad p^2
  \approx M^2 \; .
\end{align}
One can verify that no double counting occurs and that the one-loop amplitude
$\;\kreis{3}$ in \eqref{SChannelOneLoop} is cancelled by the mass counter term
$\;\kreis{2}$ up to terms of relative order $g^2\sim\Gamma\sim p^2-M^2$, as it
is required by the renormalization condition
\eqref{CMSRenormalizationCondition}.

The Laurent expansion \refeq{cutpropLaurent} is not appropriate when the
momentum $p$ is an internal loop or an integrated phase-space momentum.  Instead
we can make use of  the solutions for the decomposition $\Delta^\pm(p,\mu)$ for
small widths as given by the cutting rules \eqref{stablelimit} in
\refse{se:decomposition}.  As Veltman has shown, one does not expect unstable
particles in asymptotic states, and this dilemma is resolved at LO as follows.
Similarly to the identity \eqref{eq:identity}, we have in distributional sense
that
\begin{align}
  \Delta^+ \propto 2 \pi \theta \left(p^0 \right)\delta \left( p^2 - M^2 \right)
  \simeq  \Delta(p,\mu) 2 \Gamma M
  \Delta^*(p,\mu) \left( 1+ \mathcal{O} \left(g^2 \right) \right) = 
  \raisebox{-16.5pt}{\includegraphics{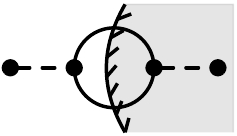}},
  \label{narrowwidth}
\end{align}
where we used \eqref{GMrelation} and \eqref{LTECutoneloopSelfEnergy}
in the last step. This shows that the LO resonant $\Delta^\pm$ can be
interpreted as a higher-order cut amplitude which is what is required
by Veltman's unitarity equation. For instance, consider the one-loop
self-energy of the stable particle denoted as $\Sigma^1_\chi$.
Computing the LTE and making use of the result \eqref{narrowwidth}
yields up to higher orders:
\begin{align}
  - 2 \Re{\Sigma^1_\chi} = 
  \raisebox{-12pt}{\includegraphics{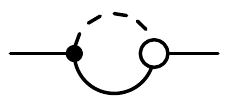}}
  =
  \begin{cases} \mathcal{O}\left( g^4 \right) & p^2 \ \text{below threshold}
  \\
  \raisebox{-14pt}{\includegraphics{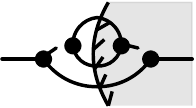}}&
  p^2 \  \text{above threshold}
  \end{cases} \;.
\label{eq:unstablecutexample}
\end{align}
This can be generalized and proves \eqref{nestedtwopoint} and
\eqref{cuttwopoint} and thus unitarity as long as the leading behaviour of
$\Delta^\pm$ is sufficient.


\subsubsection{Generalisation to higher orders}
\label{sec:generalization}
In \refse{se:resonant} we gave an interpretation for resonant $\Delta^\pm$ at
leading order in $\Gamma/M$. In the following we generalise this result to
arbitrary order by pursuing the strategy devised in steps {\strut$\;\kreis{3}$}
and $\;\kreis{4}$ in \eqref{nestedtwopoint} and \eqref{cuttwopoint}.

For LTEs of higher-order amplitudes the approximation \eqref{narrowwidth} is not
sufficient, and more terms in the expansion of $\Delta^\pm$ in $\Gamma/M$ must
be taken into account. However, with the expansion the diagrammatic
interpretation gets lost, and it is difficult to compare the result to the
right-hand side of the unitarity equation. Motivated by Veltman's approach
to derive LTEs for dressed propagators, we try to identify the LTE of
higher-order two-point functions as higher-order cut two-point functions, \ie we
aim at defining cuts that fulfil
\begin{align}
  -2 \Re{ \imag G}
  = -2\Re{
  \vcenter{\hbox{\includegraphics{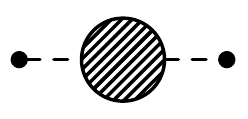}}}
  }
  \;\overset{}{=}\;
  \vcenter{\hbox{\includegraphics{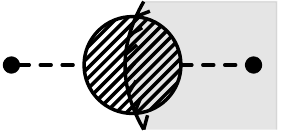}}} (-1)\;,
  \label{CutTwoPoint3}
\end{align}
where $\imag G$ represents the full propagator of the unstable
particle. Provided that \eqref{CutTwoPoint3} is true, if the LTE of
an arbitrary amplitude can be rearranged in such a way that the cut
region is given by LTEs of two-point functions [step $\;\kreis{3}$ in
\eqref{nestedtwopoint}] instead of the usual $\Delta^\pm$, then
unitarity follows immediately since we have the correct cut structure
(as shown in \refse{se:nonresonant}) and valid cuts
\eqref{CutTwoPoint3} (to be defined).  Thus, the problem is reduced to
the problem of studying LTEs of two-point functions and this is
necessary because the CMS mixes loop orders and only the LTE of the
whole two-point function can yield well-defined cuts, which we show
below.

The idea is the following: For a specific amplitude we consider all
diagrams up to a certain order in the CMS. Applying the LTE yields
contributions with the correct cut structure \eqref{ltecms}. In
contributions where unstable propagators are cut, these are
iteratively replaced by cuts of the full propagator upon including the
needed higher-order contributions [see \eqref{nestedtwopoint}]. The
validity of this replacement can be justified as follows. Consider a LTE
contribution $\tilde{\mathcal{F}}$ with a $\Delta^\pm(p,\mu)$
originating from a specific CMS propagator $\Delta(p,\mu)$ somewhere
in a diagram $\mathcal{F}$. If the diagram is of highest considered
order, we can simply use \eqref{narrowwidth} to replace
$\Delta(p,\mu)$ by a cut through stable particles. If the diagram
$\mathcal{F}$ is not of highest order, there are diagrams that have
the same structure as $\mathcal{F}$ but more self-energy insertions
next to that propagator. Among the contributions to the LTE of these
higher-order diagrams are terms that have the same structure as
$\tilde {\mathcal{F}}$, but where instead of the $\Delta^\pm$ LTE
components of two-point functions appear (originating from the
self-energy insertions), and collecting all these terms we retrieve
$\tilde{\mathcal{F}}$ with a nested LTE of a two-point function.

We first give a simple example for a LTE of an amplitude with nested two-point
functions. We show that the nested two-point functions reappear as a LTE of the
two-point functions, \ie we can identify cut two-point functions. Consider the
subset of diagrams $\tilde{\mathcal{T}}$ of the complete $2\to 2$ two-loop
amplitude $\mathcal{T}$ defined by 
\begin{align}
  \mathcal{T} \supset \tilde{\mathcal{T}}= 
  \raisebox{-19pt}{\includegraphics{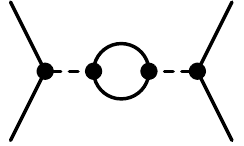}}
  +
  \raisebox{-19pt}{\includegraphics{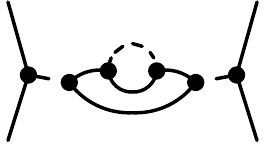}}
  =
  \raisebox{-19pt}{\includegraphics{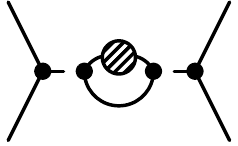}}
\;.
\label{exmaplenestedLTE}
\end{align}
In this example, the hatched circle represents the propagator of the
stable particle precisely up to one-loop order.  For the purpose of
demonstration, assume values of $s$ off the resonance which is less
complicated [the case $s\approx M^2$ is taken care of as described in
\eqref{CutTwoPoint2} below]. Computing the LTE of $\mathcal{T}$, but
only keeping the topologies of the kind of $\tilde{\mathcal{T}}$
yields
\begin{align}
  -2 \Re{\mathcal{T}}  &\supset
  \vcenter{\hbox{\includegraphics{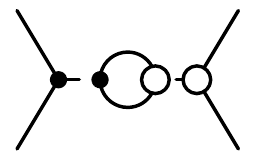}}}
  +
  \vcenter{\hbox{\includegraphics{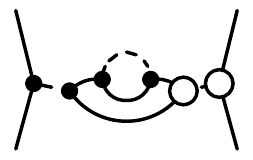}}}
  +
   \vcenter{\hbox{\includegraphics{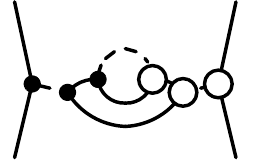}}}
  +
  \vcenter{\hbox{\includegraphics{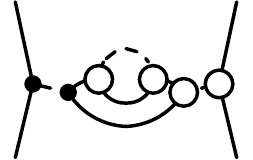}}} 
  + \mathcal{O}\left( g^8\right)\nonumber\\
   &= \vcenter{\hbox{\includegraphics{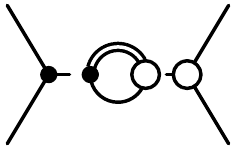}}}
  + \mathcal{O}\left( g^8\right)
  \;,
  \label{resultnestedLTE}
\end{align}
where
\begin{align}
  \raisebox{-3pt}{\includegraphics{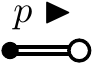}}
  :&=
  2\Re{\raisebox{-8pt}{\includegraphics{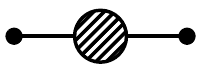}}}
  ,\qquad p^0 > 0\nonumber\\
  &=
  \raisebox{-9pt}{\includegraphics{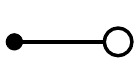}}
  +
  \raisebox{-9pt}{\includegraphics{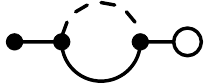}}
  +
  \raisebox{-9pt}{\includegraphics{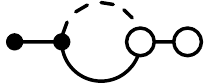}}
  +
  \raisebox{-9pt}{\includegraphics{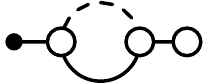}}
  \label{decompositiontwopoint}
\end{align}
represents the LTE of the nested two-point function. Underlined
endpoints in two-point functions do not come with couplings and the
underlining rules must be extended. We define the LTE of a two-point
function by pretending there were couplings $(\imag g)^2$ at the
end-points allowing us to make use of the usual underlining rules.
After removing the endpoint couplings (dividing by $g^2)$, the
difference between the two-point function with and without couplings
is a sign which is the reason why we have to take $2\Ree$ instead of
$-2\Ree$.

We arrive at
\begin{align}
  -2 \Re{\mathcal{T}} \supset
   \vcenter{\hbox{\includegraphics{CutNestedTwoPoint_part5.pdf}}}
  + \mathcal{O}\left( g^8 \right)
  =
  \begin{matrix}
    \vspace{-2.0em}
    \hspace{-0.5em}
    \raisebox{00pt}{\includegraphics[scale=0.8]{twopoint.pdf}}\\
    \raisebox{-19pt}{\includegraphics{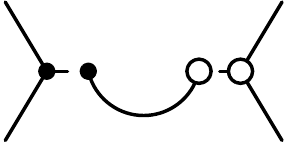}}
  \end{matrix}
  + \mathcal{O}\left( g^8 \right),
  \label{nestedLTE}
\end{align}
\ie the cut through the two-point function defined as in
\eqref{CutTwoPoint3} is expressed by a cut of a two-point function of
lower order.  After isolating cut two-point functions the normal
region $\mathcal{T}_\mathrm{L}$ and the complex-conjugated region
$\mathcal{T}^*_\mathrm{R}$ are given by
\begin{align}
  \mathcal{T}_\mathrm{L} = 
  \raisebox{-19pt}{\includegraphics{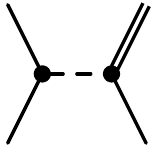}},
  \qquad \mathcal{T}^*_\mathrm{R} = 
  \raisebox{-19pt}{\includegraphics{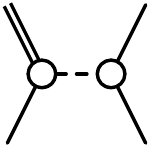}},
\end{align}
and notice that for this way of identifying terms in the LTE external
states need not to be on-shell. The reformulation of the LTE
\eqref{nestedLTE} in terms of nested LTEs of two-point functions
\eqref{decompositiontwopoint} is exactly step $\;\kreis{3}$ in
\eqref{nestedtwopoint} which we elaborate now.

In our example \eqref{exmaplenestedLTE} we derive a decomposition in
momentum space for the one-loop propagator
\begin{align}
G^1 = 
\raisebox{-0.5pt}{\includegraphics{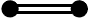}} := 
  \raisebox{-0pt}{\includegraphics{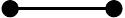}}+
  \raisebox{-1pt}{\includegraphics{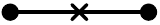}}+
  \raisebox{-9.8pt}{\includegraphics{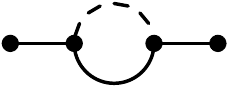}}
  \label{oneloopdecomposition}
\end{align}
in terms of $G^{1,\pm}$, to be determined, which fulfil a space--time
decomposition \eqref{decompositiontheorem} where the Feynman propagator
$\Delta_\mathrm{F}$ is replaced by $G^1$.
Such a decomposition of $G^1$ allows us to compute the LTE of
amplitudes expressed in terms of $G^1$ propagators by the sum of all
possible underlinings with suited expressions for $G^{1,+(-)}$ which
must be the same for arbitrary amplitudes, but may be different for
different propagators
(different order or different particles). Further, the LTE diagrams must
respect the cut structure since we have a physical situation and the computed
sum of LTE amplitudes must equal the sum of LTE amplitudes obtained from the
original amplitude without the identification of nested two-point functions.
In order to derive $G^{1,+(-)}$ we start from the same amplitude
\eqref{exmaplenestedLTE},
\begin{align}
  \tilde{\mathcal{T}} =
  \raisebox{-22pt}{\includegraphics{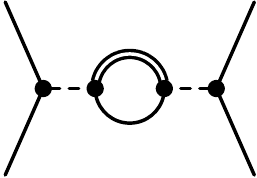}},
\end{align}
where the doubly lined propagator is defined in
\eqref{oneloopdecomposition} and computing the LTE, assuming we have a
decomposition for \eqref{oneloopdecomposition}, we obtain the result
on the left-hand side of \eqref{nestedLTE}.  The approximated
solutions for the $G^{1,\pm}$ are simply read off by comparing
with \eqref{resultnestedLTE} and are given by
$\raisebox{-1.0pt}{\includegraphics{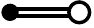}}$
\eqref{decompositiontwopoint} and
$\raisebox{-1.0pt}{\includegraphics{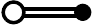}}$ the
counter part where the energy flow is in the opposite direction which
can be obtained by replacing black dots with white dots and vice versa
in \eqref{decompositiontwopoint}.

For a generic two-point function $G$ the existence of an approximate
decomposition is guaranteed by the fact that there is a decomposition
for both stable and unstable \eqref{pseudocut} tree-level propagators
and the perturbative cut structure (see \refse{se:nonresonant}), which
is a consequence of kinematics. In general, the existence of a
decomposition implies that the corresponding propagators must follow
the underlining rules in LTE diagrams and can only emerge as
unchanged, complex conjugated or cut. Moreover, the two-point function
behaves in the same way: It either stays the same, is complex
conjugated or multiple terms can be associated to the cut region where
the energy flow points in a specific direction. All other possible
outcomes violate the cut structure. The solutions $G^\pm$ for a
general two-point function are obtained by computing the LTE of $G$
and making use of the cut structure [as in the example
\eqref{decompositiontwopoint}]. The $G^{+(-)}$ is given by the
diagrams where the energy flow is to the right(left).

Returning to \eqref{nestedtwopoint}, instead of computing the LTE of
an amplitude which is given by vertices and tree-level propagators, we
can think of the same amplitude, but reformulated in terms of
$\raisebox{-0.5pt}{\includegraphics{twopointdecomp.pdf}}$. In
the LTE we encounter cut propagators like
\eqref{decompositiontwopoint} which can be substituted by the real
part of the two-point function, leading exactly to the right-hand side of
\eqref{nestedtwopoint}.

Having shown \eqref{nestedtwopoint}, \ie how to rearrange LTEs of
matrix elements in order to identify nested LTEs of two-point
functions, we turn to the proof of \eqref{cuttwopoint} and elaborate
on the meaning of \eqref{CutTwoPoint3}.  
The diagrammatic significance of $\Gamma M$ turns out to be a real
problem beyond one loop in our current framework.
At some point in our calculation, in particular when $\Delta^\pm$ is
resonant, we have to plug in the expression for $\Gamma M$ obtained
from the renormalization condition.  This problem can be circumvented
by making use of resummed results, \ie instead of using the usual
perturbative expansion we represent two-point functions by their fully
resummed equivalent deliberately taking into account non-significant
(higher) perturbative orders.  Then, the partial resummation of the
CMS is replaced by a complete resummation which turns out to be
sufficient for a diagrammatic interpretation. Returning to the
statement that, in contrast to the on-shell scheme, $\Gamma M$ does
not represent well-defined cut-contributions, one realises that
\begin{align}
  2 \Re{ \imag \tilde \Sigma_{\mathrm{R}}(p^2)}:=  
  2 \Re{\imag \Sigma_{\mathrm{R}}(p^2)-\Gamma M} =
  2 \Re{\imag (\Sigma_{\mathrm{R}}(p^2)-\mu^2)}
  \label{selfenergycut}
\end{align}
does, which can be understood as follows. We express the fully
resummed two-point function as
\begin{equation}
  \frac{\imag}{p^2-\mu^2 +  \Sigma_{\mathrm{R}} } =
  \frac{\imag}{p^2-\mu^2  +  \Sigma_{\mathrm{R}} }
  \left(\frac{p^2-\mu^2 +  \Sigma_{\mathrm{R}} }{\imag}\right)^{*}
  \left(\frac{\imag}{p^2-\mu^2 +  \Sigma_{\mathrm{R}} }\right)^{*},
  \label{CutTwoPoint}
\end{equation}
and computing the LTE of this expression we obtain
\begin{equation}
  -2\Re{\raisebox{-14.5pt}{\includegraphics{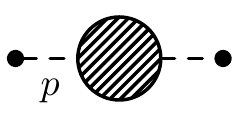}}}
  =
  \raisebox{-14pt}{\includegraphics{CutTwoPoint_part2.pdf}}
  2 \Re{\raisebox{-11pt}{\includegraphics{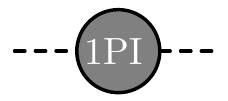}}}
  \raisebox{-17pt}{\includegraphics{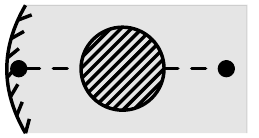}}
  \specialnumber{a}
  \label{CutTwoPoint2}
\end{equation}
\begin{equation}
  \overset{!}{=}
  \raisebox{-16.8pt}{\includegraphics{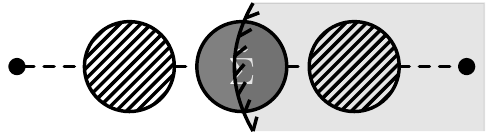}} (-1)\,.
  \specialnumber{b}
  \tag{\ref{CutTwoPoint2}}
\end{equation}
Note that when taking the real part of \eqref{CutTwoPoint} one must only compute
the real part of $\bigl[\imag \left(p^2-M^2\right)+\Gamma M -
(\imag\Sigma_{\mathrm{R}})^{*}\bigl]$ because the other factors form a real
number. Further, the step \specialeqref{CutTwoPoint2}{a} is only allowed if
${\imag}/(p^2-\mu^2 + \Sigma_{\mathrm{R}})$ is non-singular which is true for
unstable particles. Equation \specialeqref{CutTwoPoint2}{a} expresses the LTE
of two-point functions by the LTE of self-energies. The equality with
\specialeqref{CutTwoPoint2}{b} is equivalent to \eqref{CutTwoPoint3} allowing us
to properly define cut two-point functions. Expanding the full propagators in
\specialeqref{CutTwoPoint2}{b} in the CMS implies that the right-hand side of
\eqref{CutTwoPoint3} should be defined as
\begin{align}
  \vcenter{\hbox{\includegraphics{CutTwoPoint3_part1.pdf}}}
  := 
  \raisebox{-17.8pt}{\includegraphics{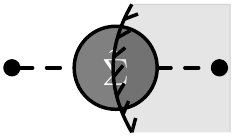}}
  +
  \raisebox{-16.8pt}{\includegraphics{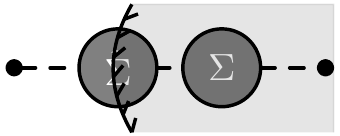}}
  +
  \raisebox{-16.8pt}{\includegraphics{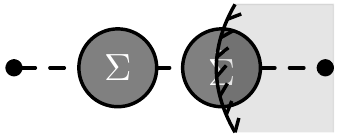}}
  + \ldots\,,
  \label{CutTwoPoint4}
\end{align}
where the cut self-energies are defined by \eqref{selfenergycut} and normal and
complex-conjugated self-energies are just the usual ones renormalized according
to the CMS. This definition of cuts is in agreement with Veltman's unitarity
equation \eqref{veltmanunitarity} as we show in the sequel.

In \eqref{selfenergycut}, \eqref{CutTwoPoint},
\specialeqref{CutTwoPoint2}{a} and \specialeqref{CutTwoPoint2}{b}, \ie
for dressed propagators there is no partial resummation. Starting from
the CMS, we can construct dressed propagators by resumming
renormalized self-energies. After full resummation the $\imag \Gamma
M$ in the propagator cancels with the $\imag \Gamma M$ of $\imag
\Sigma_\mathrm{R}$ and all explicit $\imag \Gamma M$ expressions
disappear which is the reason why \eqref{selfenergycut} represents
well-defined cuts. In this limit unitarity is not violated as has been
shown by Veltman \cite{Veltman:1963} and it is left to understand that
nothing goes wrong when going from full resummation to the CMS.  This
is formally shown as follows: We only need to study cut two-point
functions, \ie we need to show that the right-hand side of
\specialeqref{CutTwoPoint2}{a} is equal to
\specialeqref{CutTwoPoint2}{b}.  Assume the left-hand side of
\specialeqref{CutTwoPoint2}{a} is given at $n$-loop order, then
\specialeqref{CutTwoPoint2}{a} tells us that the LTE of an $n$-loop
two-point function can be computed by LTEs of $n$-loop self-energies,
where we actually mean the self-energies \eqref{selfenergycut}. As in
our example \eqref{nestedLTE}, these self-energy LTEs are iterated
LTEs of one- to $(n-1)$-loop two-point functions.  Thus, one makes the
induction hypothesis that LTEs of two-point functions
\eqref{CutTwoPoint2} represent well-defined cuts at $n-1$ loops.
Expanding the two-point function on the left-hand side of
\specialeqref{CutTwoPoint2}{a} at $n$ loops the statement
\specialeqref{CutTwoPoint2}{b} follows from the induction hypothesis,
the start of the induction being given by
\eqref{LTECutoneloopSelfEnergy}. Thus, cuts are defined iteratively as
in the example \eqref{nestedLTE} and each cut through an unstable
particle, possibly belonging to higher-order contributions and after
collecting all terms like in \eqref{eq:unstablecutexample} results in
a nested cut two-point function \eqref{CutTwoPoint4} which itself can
have cut unstable particles.

Let us illustrate the procedure at the example of our toy theory. Consider the
two-point function of an unstable particle at two-loop order
\begin{align}
  \imag G^2_\phi = 
  \raisebox{-12.0pt}{\includegraphics{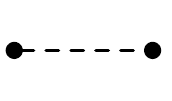}}
  +
  \raisebox{-12.0pt}{\includegraphics{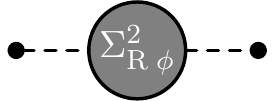}}
  +
  \raisebox{-12.0pt}{\includegraphics{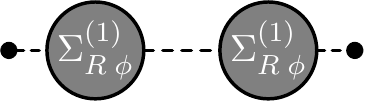}}\,,
  \label{amptwoloop}
\end{align}
\noindent
where $\Sigma^{2}_{\mathrm{R},\phi}=\Sigma^{(1)}_{\mathrm{R},\phi}+
\Sigma^{(2)}_{\mathrm{R},\phi}$ is the two-loop self-energy of the
unstable particle $\phi$ renormalized according to the CMS, and
$\Sigma^{(1)}_{\mathrm{R},\phi}$ and
$\Sigma^{(2)}_{\mathrm{R},\phi}$ denote the one-loop and two-loop
renormalized contributions to the two-loop self-energy, respectively.
At this point we have to demand that the perturbation series is valid
in the sense that the two-point function is well approximated by
\begin{align}
  \imag G^2_\phi = \frac{\imag}{p^2-M^2 + \imag \tilde
  \Sigma_{\mathrm{R}, \phi}^2 } \left(1 + \mathcal{O}\left(g^6\right)\right)=: 
  \raisebox{-10pt}{\includegraphics{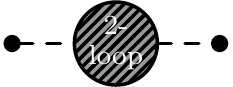}}
  \left(1 + \mathcal{O}\left(g^6\right) \right),
\end{align}
where the two-point function on the right-hand side is defined as the resummed
propagator for which the self-energy is evaluated and renormalized at two-loop
order. Then, we can compute the LTE with the help of
\specialeqref{CutTwoPoint2}{a} and up to higher orders we have
\begin{align}
  -2 \Re{\imag {G}^2_\phi} = 
  \raisebox{-9.5pt}{\includegraphics{CutTwoPoint_part7.pdf}}
  \;2\Re{
  \raisebox{-15.5pt}{\includegraphics{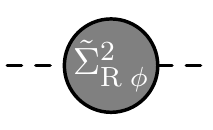}}}
  \raisebox{-16.5pt}{\includegraphics{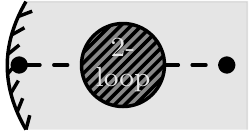}},
  \label{CutTwoPointTwoLoop}
\end{align}
where $\tilde \Sigma_{\mathrm{R}, \phi}^2$ is given by
\begin{align}
  \tilde \Sigma_{\mathrm{R}, \phi}^2 =  \Sigma^{(2)}_\phi + \Sigma^{(1)}_\phi
  - \Re{ \delta \mu^2}.
\end{align}
Notice that the imaginary part of $\delta \mu^2$ dropped out and only the real
part is left over renormalising the one- and two-loop self-energies. In what
follows we could have argued with the induction hypothesis, but we work out this
example explicitly and compute the LTE of $\Sigma^{(2)}_\phi$ and $
\Sigma^{(1)}_\phi$. The LTE of $\Sigma^{(1)}_\phi$ is trivial and yields the
one-loop cut \eqref{LTECutoneloopSelfEnergy}. Among all contributions to
$\Sigma^{(2)}_\phi$ there is none with nested unstable two-point functions
except for the CMS propagator. For instance, consider the example
\begin{align}
  \Sigma_\phi^{(2)} \supset
  \raisebox{-14pt}{\includegraphics{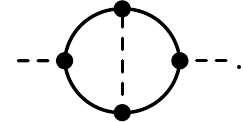}}
  \label{TwoLoopSelfEnergyExample}
\end{align}
We apply the LTE and we only keep the terms with the correct cut structure since
the other ones are negligible. Then we can compute the LTE for the two-point
functions or matrix elements with the help of \eqref{CutTwoPointTwoLoop}.
Keeping only the terms directly related to our example
\eqref{TwoLoopSelfEnergyExample}, we obtain
\begin{align}
  \raisebox{-25pt}
  {\includegraphics{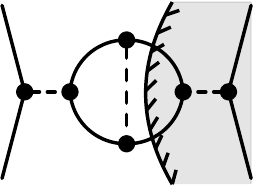}}
  +
  \raisebox{-25pt}
  {\includegraphics{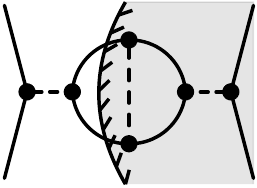}}
  +
  \raisebox{-25pt}
  {\includegraphics{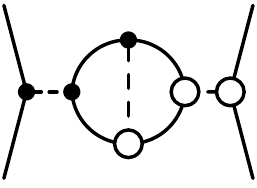}}
  +
  \raisebox{-25pt}
  {\includegraphics{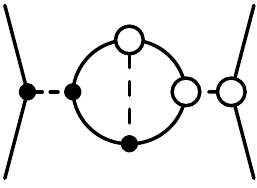}}.
\end{align}
As already mentioned, since there are no nested unstable two-point functions, we
can directly apply \eqref{narrowwidth}
\begin{align}
  \raisebox{-25pt}
  {\includegraphics{unstabletwoloopselfenergycut_part3.pdf}}
  =
  \raisebox{-25pt}
  {\includegraphics{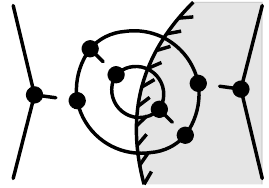}}
  \left( 1 + \mathcal{O}\left(g^2\right)\right).
\end{align}
In this way we can show that all two-loop self-energies in this theory yield, up
to higher orders, well-defined cut self-energies. Having dealt with the two-loop
self-energies, if we combine the one- and two-loop result in
\eqref{CutTwoPointTwoLoop}, we have explicitly shown
\specialeqref{CutTwoPoint2}{b} at two-loop order. In case of nested unstable
two-point functions we would have had to use the induction hypothesis.

Finally, our results can be summarised as follows:
\begin{itemize} 
\item In the beginning we demonstrated how to compute LTEs of arbitrary
  amplitudes leading us to the result that the cut structure is guaranteed in a
  perturbative sense which is the basis for unitarity.  One is left to check
  that $\Delta^\pm$ yields well-defined cuts.
  
\item From the solutions of $\Delta^\pm$ for CMS propagators we concluded that
  unitarity holds off resonances. The leading behaviour of resonant $\Delta^\pm$
  can be interpreted as the cut one-loop two-point function [see
  \eqref{narrowwidth}].

\item The leading approximation of $\Delta^\pm$ is not enough beyond one-loop,
  and higher-order corrections need to be included. The LTEs of different loop
  orders do not separately represent valid cuts in the sense of Veltman's
  definition and must be considered simultaneously because owing to the
  partial resummation the diagrams are connected to each other and
  cancellations take place. Whenever there is a resonant unstable $\Delta^\pm$
  we identify cuts by iteratively including appropriate higher-order
  contributions resulting in nested LTEs of two-point functions. For the LTE
  of the two-point functions up to a given order a valid cut interpretation
  can be assigned which is consistent with the interpretation of Veltman, \ie
  only lines of stable particles are cut.

\end{itemize}


\section{Conclusions}
\label{conclusions}
The Complex-Mass Scheme provides a straightforward method to consistently
implement unstable particles in perturbative calculations. Formally, the
procedure is an analytic continuation of matrix elements to complex masses and
(if necessary) couplings with appropriate renormalization condition.

In the Complex-Mass Scheme the Cutkosky cutting rules can no longer be used to
verify unitarity, and it was not clear how perturbative unitarity is
implemented.  Following Veltman, we derived a Largest-Time Equation within the
Complex-Mass Scheme which could then be used to obtain a diagrammatic
representation for the imaginary part of scattering amplitudes, also when
unstable particles are present.

Our derivation of the Largest-Time Equation is based on the decomposition
theorem and we showed that an appropriate decomposition can be achieved for the
Complex-Mass Scheme propagator. As a result, one finds that the would-be cut
propagators $\Delta^\pm(p,\mu)$ of unstable particles are smoothed versions of
the stable ones. In case of stable particles the Largest-Time Equation coincides
with the Cutkosky cutting rules, but including unstable particles leads to
additional contributions which can be interpreted as contributions where the
energy flow is backward. Performing an expansion solely of would-be cut
propagators $\Delta^\pm(p,\mu)$ of unstable particles in $\Gamma / M$ does
indeed yield cutting rules where unstable resonant $\Delta^\pm(p,\mu)$ can be
replaced by higher-order cuts through stable particles only. In this way, we
recover the perturbative statement of Veltman's result in the Complex-Mass
Scheme, namely that a QFT is unitary up to higher orders only if unstable
particles are excluded from asymptotic states. While we only considered a toy
model with real couplings, the generalisation to complex couplings is
straightforward.

\end{document}